\documentclass[useAMS,usenatbib,usegraphicx,twocolumn]{mn2e}
  
\usepackage{float}
\usepackage{aas_macros}
\usepackage{threeparttable}
\usepackage{grffile}
\usepackage{amsmath, amssymb}
\usepackage{lastpage}
\usepackage{multirow}
\usepackage{wallpaper}
\usepackage{mathbbol}
\usepackage{rotating}
\usepackage{blindtext}
\usepackage[T1]{fontenc}
\usepackage{changes}
\usepackage{lscape}

\usepackage{tocloft}
\newcommand{\ita}{\textit}

\newcommand{\beq}{\begin{equation}}
\newcommand{\eeq}{\end{equation}}  
\newcommand{\RNum}[1]{\uppercase\expandafter{\romannumeral #1\relax}}

  \title[Turbulence in MCs]{The turbulence driving parameter of molecular clouds in disc galaxies}
  
  \author[B.~K\"ortgen ]{Bastian~K\"ortgen\\
  Hamburger Sternwarte, Universit\"at Hamburg, Gojenbergsweg 112, D-21029 Hamburg, Germany \\
  }

\date{Released 2020}

\begin{document}

\label{firstpage}
\maketitle

\begin{abstract}
Supersonic turbulence plays a pivotal role during the formation of molecular clouds and stars in galaxies. 
However, little is known about how the fraction of compressive and solenoidal modes in the velocity field evolves over time and how it depends on properties of the molecular cloud or 
the galactic environment. In this work, we carry out magnetohydrodynamical simulations of disc galaxies and study the time evolution of 
the turbulence driving parameter for an ensemble of clouds. We find that the time-averaged turbulence driving parameter is 
insensitive to 
the position of the cloud within the galaxy. The ensemble-averaged driving parameter is found to be rather compressive with 
$b\sim0.5-0.7$, indicating almost time-independent global star formation properties. However, each individual cloud shows a highly fluctuating driving parameter, which would strongly affect the 
cloud's star formation rate. We find that the mode of 
turbulence driving can rapidly change within only a few Myr, both from solenoidal to compressive and vice versa. We attribute 
these changes to cloud collisions and to tidal interactions with clouds or overdensities in the 
environment. Last, we find no significant differences in the average driving parameter between hydrodynamic and initially strongly 
magnetised galaxies. However, the magnetic field tends to reduce the overall fluctuation of the driving parameter. The average 
driving as well as its uncertainty are seen to be in agreement with recent constraints on the turbulence driving mode for 
solar neighbourhood clouds.
\end{abstract}
\begin{keywords}
galaxies: ISM; ISM: kinematics and dynamics; ISM: magnetic fields; ISM: clouds; stars: formation
\end{keywords}

\section{Introduction}
Stars are born within self-gravitating, magnetised and turbulent molecular clouds. The complex 
dynamical interplay between gravity, turbulence, magnetic fields and 
(at more evolved stages) stellar feedback strongly shapes the density distribution of the 
parental cloud and thus determines the rate and efficiency at which stars 
form \citep{MacLow04,Elmegreen04,Scalo04,Federrath12}. 
For supersonically 
turbulent, isothermal (and magnetised) gas the density probability distribution function (PDF) resembles 
a lognormal distribution, which means that the logarithm of the density (contrast) is Gaussian 
distributed \citep{Vazquez94,Passot98,Federrath08}. However, the 
interstellar medium (ISM) of galaxies is a multiphase gas, strongly influenced by gravitational forces. 
Hence, the density PDF will take several forms that indicate, which physical process is dominant
or at least relevant \citep{Kainulainen09,Ballesteros11a,Schneider14}. For a thermally bistable gas, the density PDF becomes 
double-peaked. Each peak corresponds to a single thermal phase of the gas, and each phase can be fitted 
by a lognormal distribution \citep{Seifried11c}. 
When the gas has become sufficiently dense, gravity starts to become 
dominant. The subsequent flow of gas towards smaller scales and the resultant formation of power-law 
density profiles induces the 
formation of a power-law tail in the density PDF at high densities \citep{Kritsuk11a}.  Over time, the transition point from the 
log-normal to the power-law part in the PDF will shift towards lower densities, as gravity starts to impact the 
more diffuse gas in the cloud \citep{Burkhart17}. In addition, the power-law tail flattens with increasing star formation 
activity of the cloud \citep{Federrath13}. However, in case the bistable gas is subjected to stellar feedback or 
strong turbulent phase mixing, a power-law tail can also form in a non-gravitating medium, thereby complicating the 
interpretation of the density PDF \citep{Seifried11c,Tremblin14}. \\
An essential ingredient for current theoretical models of star 
formation, based on turbulent fragmentation of the gas, is the width of the density distribution 
\citep{Krumholz05c,Padoan11a,Hennebelle11,Federrath12,Zamora12,Konstandin16,Voelschow17}. Theoretical 
studies have shown that there exists a relation between the variance of the density PDF and gas-dynamical 
quantities, such as the turbulent Mach number, $\mathcal{M}$, the ratio of thermal to magnetic pressure, $\beta$ 
and the fraction of compressive or solenoidal modes in the 
turbulent velocity field, typically denoted by $b$. For a fully compressive driving of the turbulent field $b=1$, whereas purely 
solenoidal driving implies $b=1/3$ \citep[see][for a first detailled investigation]{Federrath10b}. Therefore, the gas dynamics, which affect the density variance, implicitly 
affect the star formation properties of the cloud. For example, \citet{Federrath12} have shown that 
the star formation rate of a turbulent molecular cloud with equal Mach number is increased by a factor of $\gtrsim10$, when the turbulent velocity field is changed from solenoidal to compressive. \\
Rather than treating the driving parameter, $b$, as an input parameter, several recent studies focused on 
retrieving it from the self-consistent gas dynamics in the analysed environment \citep{Koertgen17b,Jin17,Menon20}. 
While \citet{Koertgen17b} and \citet{Jin17} find $b$ to vary between $b\sim0.3$ and $b\sim0.8$ in their different 
simulations of molecular cloud formation, the study by \citet{Menon20} finds a rather stable $b\sim0.76$ in a 
simulated region, which is affected by radiative feedback. Furthermore, \citet{Pan16} argue that 
supernova feedback induces mostly solenoidal motions due to the onset of the baroclinic instability in the post-shock region of the supernova remnant. These results all emphasise the 
impact of the various feedback mechanisms and involved physics on the turbulent velocity field.\\
Deriving the turbulence driving parameter from observations of molecular clouds is even more challenging. This 
is mostly due to the restriction of the data to at most two dimensions. However, it is nevertheless useful and 
necessary to confront theory with reality. For molecular clouds in the solar 
neighbourhood, \citet{Ginsburg13} derived a lower limit of $b\sim0.4$. \citet{Orkisz17} analysed the solenoidal 
fraction of the velocity field in the Orion~B cloud and showed that the fraction varies both as a function of 
position across the complex and radial distance to a given center (e.g. the position of the peak density). In 
terms of the driving parameter, this latter study deduced that $b$ varies from $b>0.3$ to $b\lesssim1$, i.e. 
the gas motion is neither fully solenoidal nor entirely compressive. Their most likely value was about $b\sim0.4$, 
which indicates equipartition between the two modes \citep{Federrath10b}. \citet{Kainulainen17} studied a set of 13 solar 
neighbourhood clouds and argued that $b\in\left[0.4,0.7\right]$ provides a strict range of values. As these 
authors point out, the driving parameter was derived from the density variance - Mach number relation, despite the 
fact that their data did not necessarily follow such a relation. In contrast to clouds in the solar neighbourhood, the 
Galactic center may provide a significantly different environment. As \citet{Federrath16d} show in their 
study of the central molecular zone cloud G0.253+0.016, the 
driving of turbulence in this region might be almost fully solenoidal with $b\sim0.22\pm0.12$ due to the 
increased shear. Hence, understanding the 
driving of turbulence in different environments of the Galaxy and within molecular clouds is crucial for the 
understanding of the local gas dynamics and the subsequent star formation process.\\
This work strives to add further insights into the driving of turbulence by analysing the time evolution of the 
turbulence driving parameter in a set of molecular clouds formed in disc galaxies. The study is structured as 
follows: In Sect.~\ref{sectIC} we briefly discuss the used numerical tools and initial conditions. Sect.~\ref{sect:results} discusses the findings of this study, before it is closed with a summary in 
Sect.~\ref{sect:summary}.
\section{Numerical Method and Initial Conditions}
\label{sectIC}
\subsection{Methods}
We use the finite volume code \textsc{flash} \citep[version 4.2.2,][]{Dubey08}. The equations of ideal 
magnetohydrodynamics are solved each timestep on the numerical grid with a five-wave Riemann solver \citep{Bouchut09,Waagan11}. To ensure the solenoidal 
constraint of the magnetic field, we use a hyperbolic cleaning scheme \citep{Dedner02}. Poisson's equation 
for the self-gravity of the gas is solved with a Barnes-Hut tree solver \citep[see e.g.][for a GPU version]{Lukat16}. 
The thermodynamic evolution of the gas is treated by optically thin heating and cooling, where we follow the 
approach by \citet[][with modifications by \cite{Vazquez07}]{Koyama02}, i.e. the heating rate is constant and the 
temperature-dependent cooling rate is provided in tabulated form.\\
The cubic simulation domain has an edge length of \mbox{$L=40\,\mathrm{kpc}$} and the root grid is at a grid resolution of 
\mbox{$\Delta x_\mathrm{root}=625\,\mathrm{pc}$}. Over the course of the evolution, the grid is adaptively refined
\citep[][]{Berger84} once the local Jeans length is resolved with less than 32 grid cells \citep[as recommended based on the results in][]{Federrath11a}. De-refinement is 
initiated when the Jeans length instead is resolved by more than 64 grid cells. In total, we allow for a maximum of 
11 refinement levels, which gives a peak spatial resolution of $\Delta x_\mathrm{peak}=19.5\,\mathrm{pc}$. The gas scale height is thus resolved with $\gtrsim10-15$ grid 
cells in the major part of the disc \citep{Koertgen19b}. Any fragmentation at later times from $R\sim4-5\,\mathrm{kpc}$ 
on can thus assumed to be physical \citep[see also][]{Truelove97}. To extract 
gas dynamics within the formed clouds solely driven by gravity,
neither stellar feedback nor sink particles are included in the current study and we introduce an artificial pressure 
term on the highest level of refinement.

\subsection{Initial Conditions}
An in depth description of the initial conditions is given in \citet[][see also \cite{Koertgen18L}]{Koertgen19b}, so we provide only a brief overview here.\\
We set up a thin disc, following the approach by \citet{Tasker09}, with a radially increasing 
gas scale height. The gas density profile is adjusted in such way that the Toomre stability 
parameter is initially constant at $Q=2$ in the major part of the disc. In the inner most part it is 
set to $Q=20$ to increase the stability of the less well resolved region. The final mass of the disc 
is $M_\mathrm{disc}\approx10^{10}\,\mathrm{M}_\odot$, similar to the mass of the LMC. We use a fixed external 
logarithmic gravitational potential to account for the effect of dark matter and old stars, which also provides 
a flat rotation curve. The majority of the performed simulations starts with a purely toroidal 
magnetic field, where the field strength is coupled to the gas density, i.e. $B\propto\varrho^{1/2}$, 
which results in a constant ratio of thermal to magnetic pressure. We do not impose any turbulent 
perturbation in order to simplify the initial conditions. Hence, any local velocity 
fluctuations at later stages are the result of gravitational fragmentation and self-consistently 
generated dynamics (i.e. cloud-cloud interactions).

\section{Results}\label{sect:results}
\subsection{Global view in a nutshell}
\begin{figure*}
    \centering
        \includegraphics[width=0.3\textwidth]{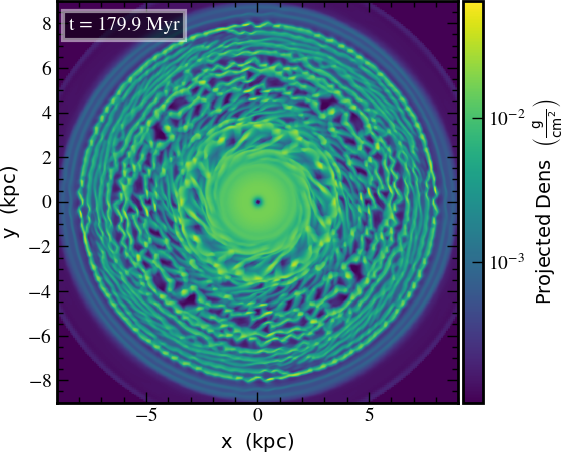}
        \includegraphics[width=0.3\textwidth]{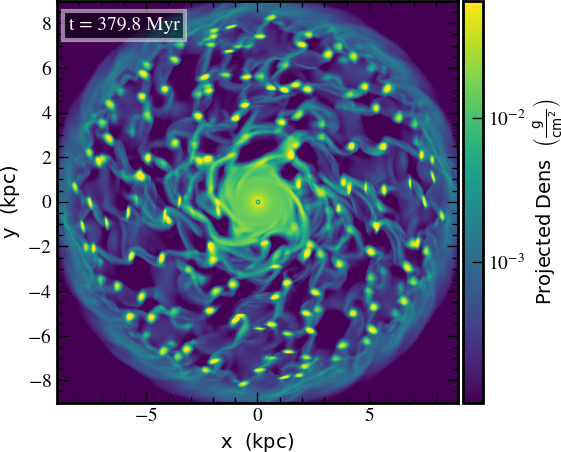}
        \includegraphics[width=0.3\textwidth]{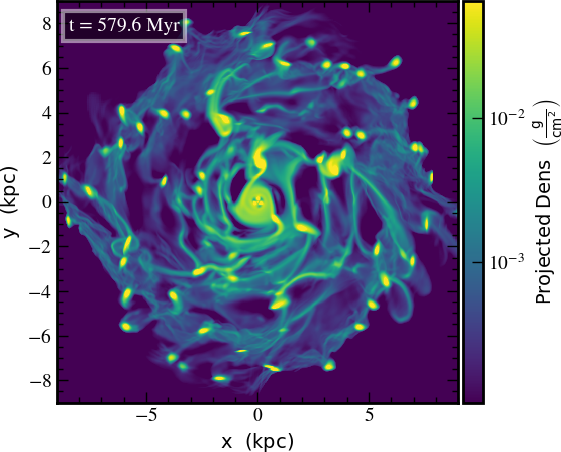}\\
        \includegraphics[width=0.3\textwidth]{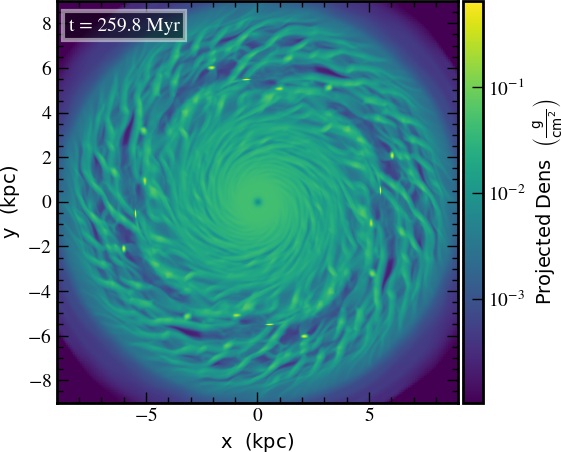}
        \includegraphics[width=0.3\textwidth]{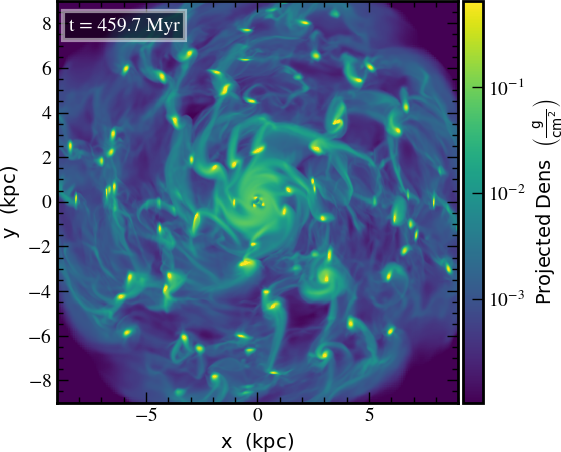}
        \includegraphics[width=0.3\textwidth]{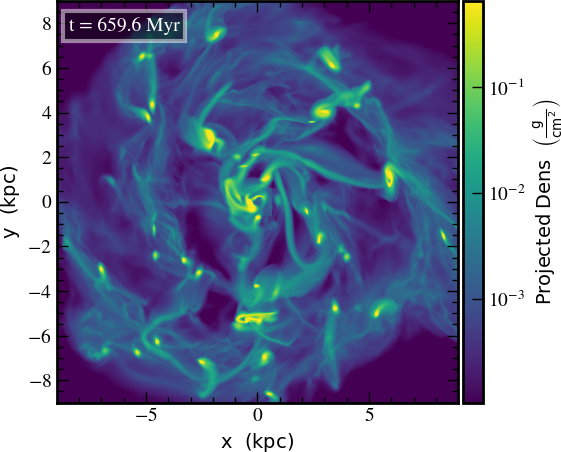}
        \caption{Surface density maps of the hydrodynamic (top) and magnetised disc (bottom) at three 
        times, starting from the point where the first diffuse ($n_\mathrm{min}\geq10\,\mathrm{cm}^{-3}$) clouds have been identified. Note the 
        different morphology of the fragmented disc and the increased density in the diffuse gas 
        for the MHD disc at later times.}
        \label{fig:glovie_cdens}
\end{figure*}

\begin{figure*}
    \centering
        \includegraphics[width=0.4\textwidth]{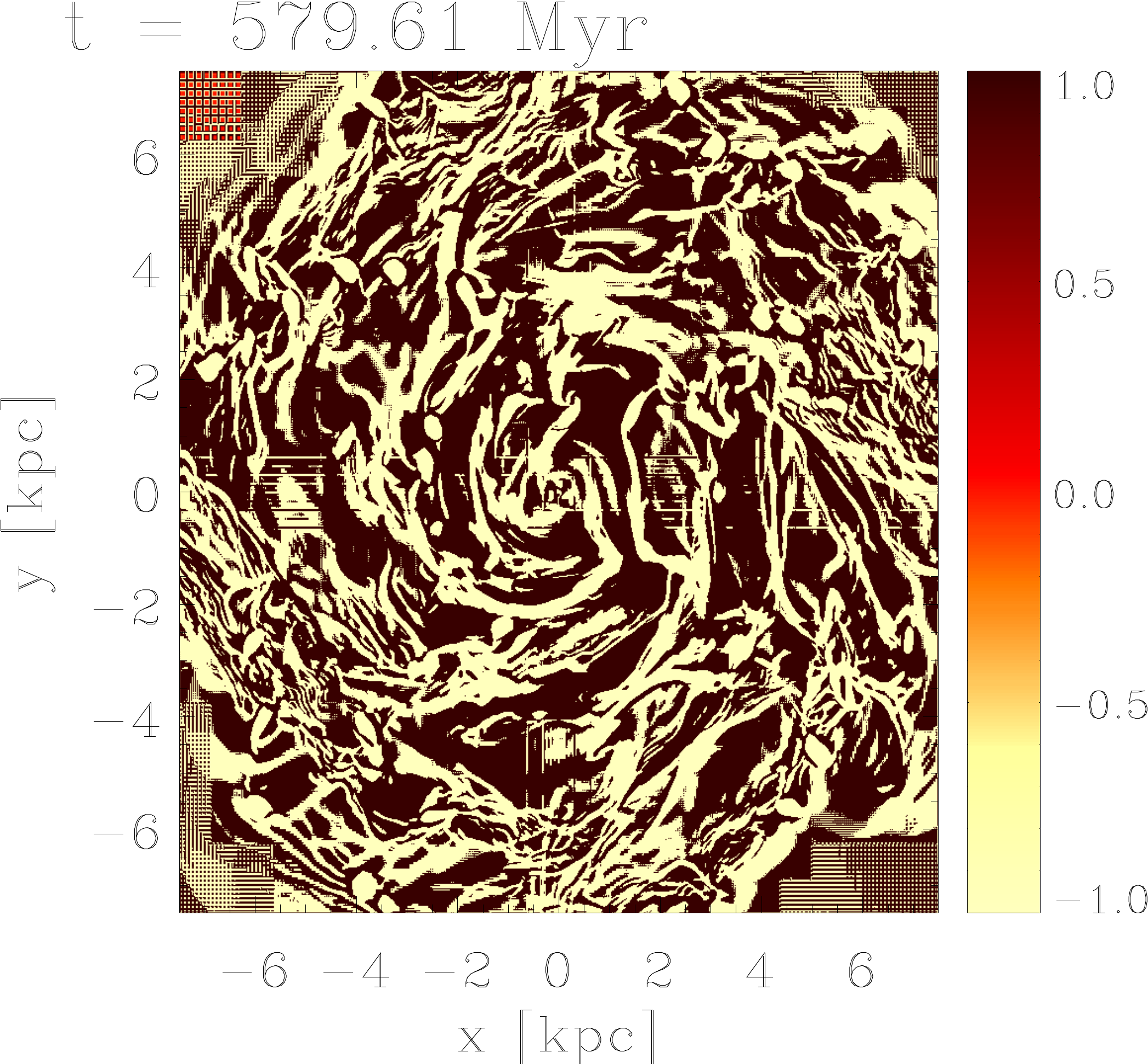}
        \includegraphics[width=0.4\textwidth]{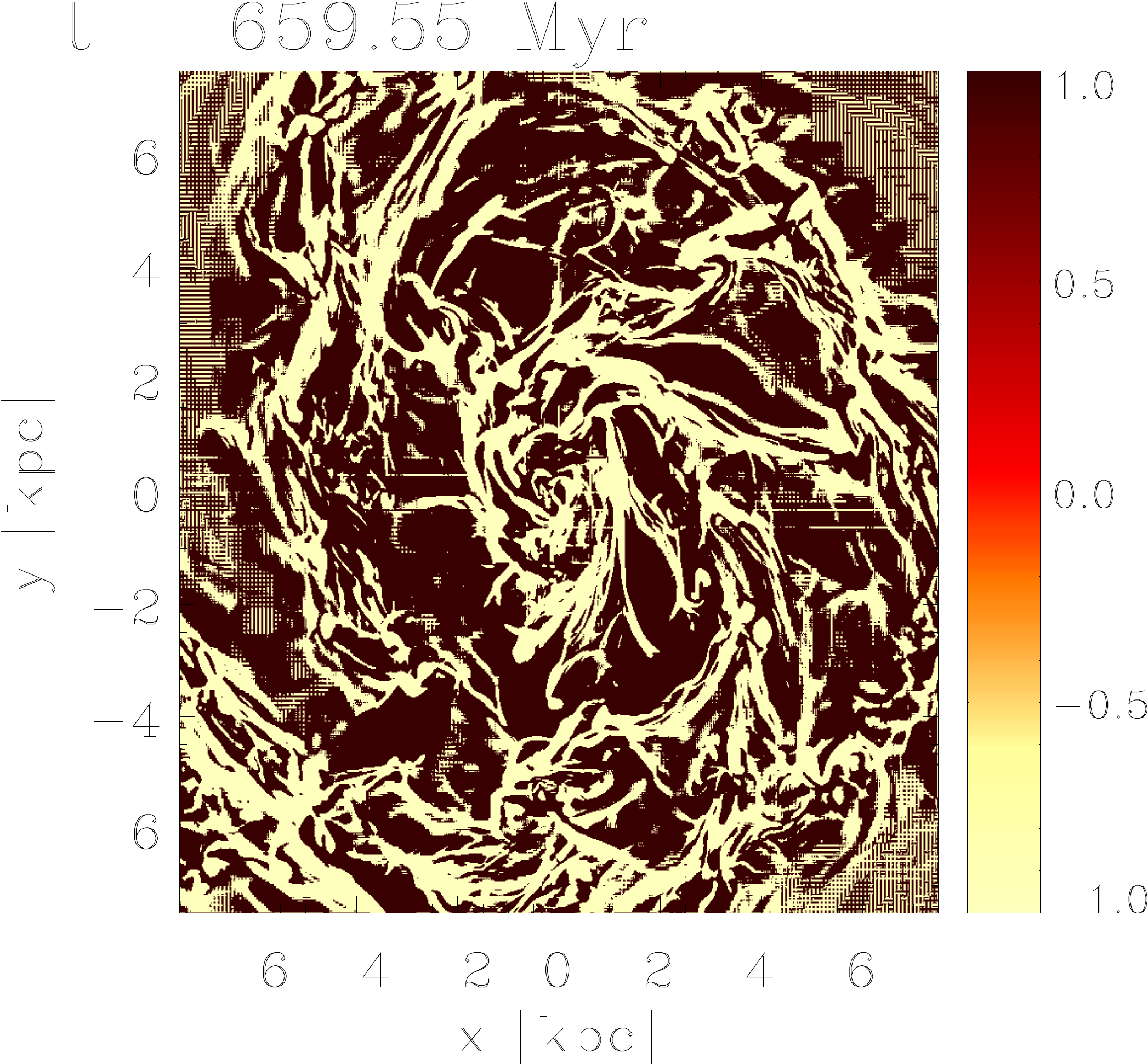}\\
        \includegraphics[width=0.4\textwidth]{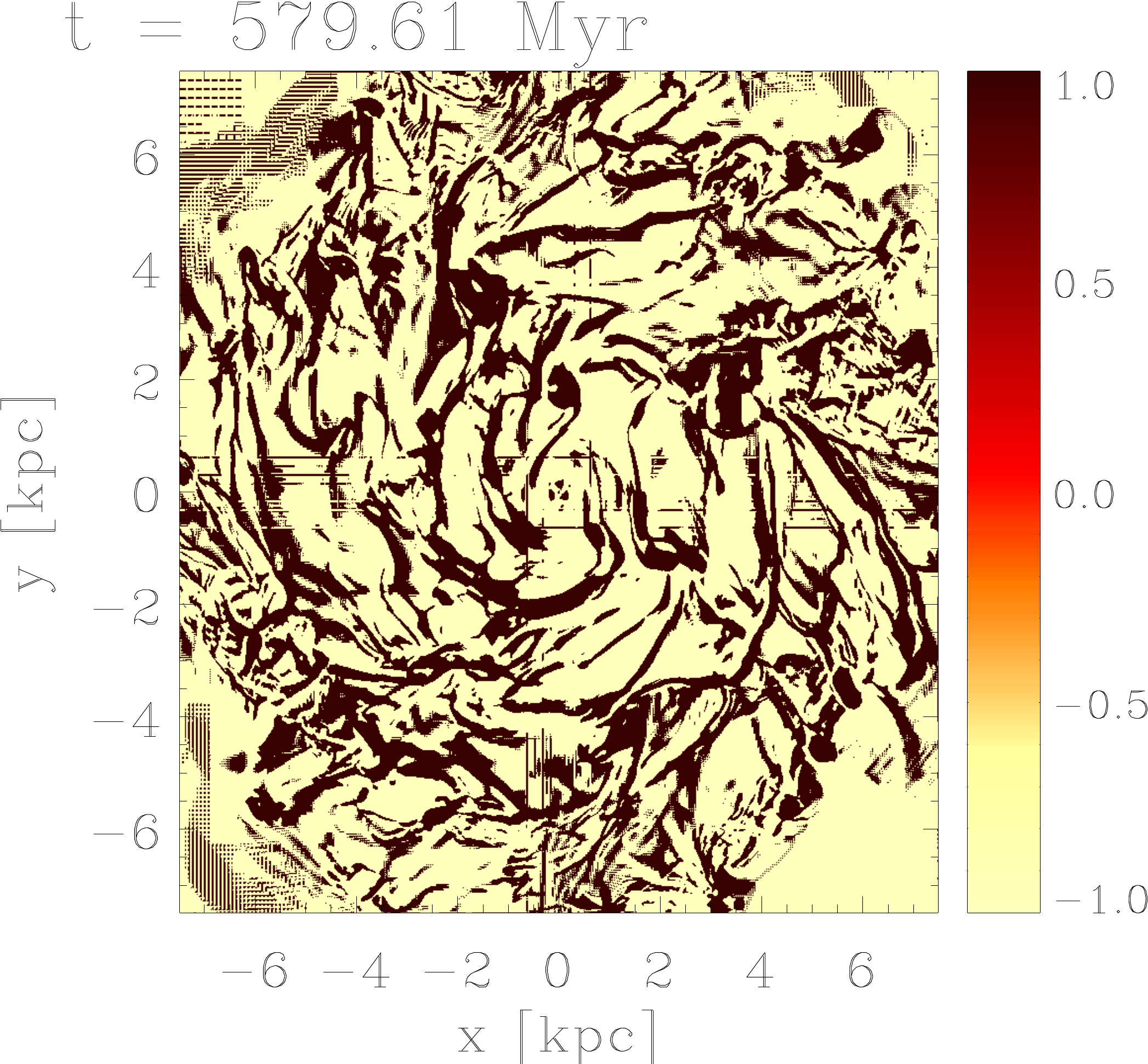}
        \includegraphics[width=0.4\textwidth]{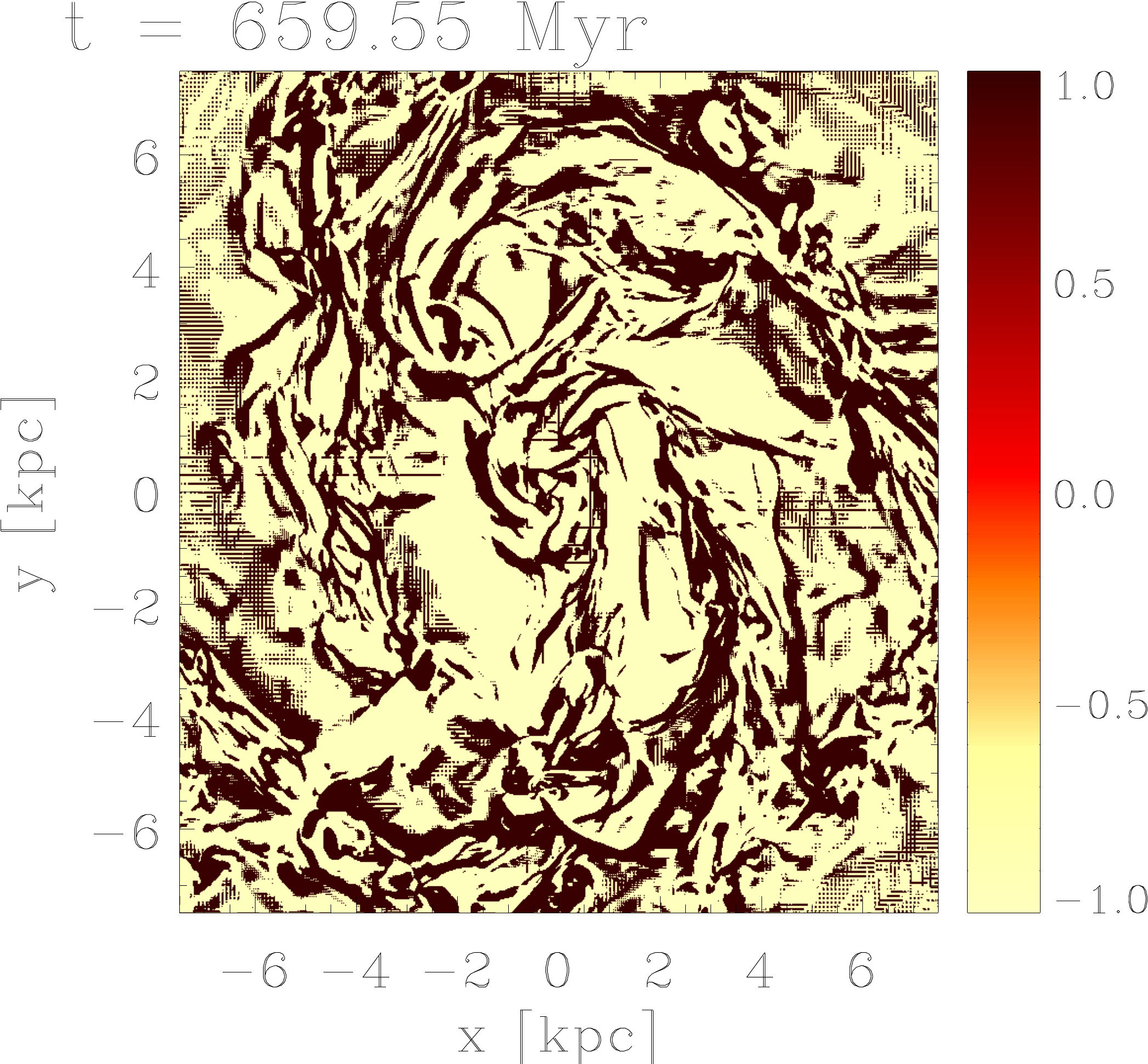}
        \caption{\ita{Top:} Divergence of the projected residual (rotation curve 
        subtracted) velocity field for the hydrodynamic (left) and magnetised (right) 
        discs shortly before the end of the simulation. Both discs reveal filamentary structures of the divergence field. However, there appear more converging regions in the hydrodynamic disc. Note the large patches of 
        diverging gas in the MHD disc. \ita{Bottom:} Vorticity. The colour bar is arranged that converging 
        regions are shown in bright colour, while diverging regions are shown as the 
        dark patches. In the bottom panel, the same colour coding holds for negative 
        and positive vorticity.  }
        \label{fig:glovie_vdisp}
\end{figure*}

Fig.~\ref{fig:glovie_cdens} presents column density maps of the hydrodynamic (top panels) and the 
magnetised (bottom) discs, separated in time by $\Delta t=200\,\mathrm{Myr}$, which corresponds to 
almost an entire disc rotation at $R=8\,\mathrm{kpc}$.\\
With time, the initially smooth density profile is distorted, because the discs fragment into 
individual overdensities. While the hydrodynamic disc breaks up into several rings due to the onset 
of the Toomre instability, the magnetised galaxy builds up filamentary structures, which extend into 
the radial and azimuthal directions. The latter fragmentation pattern is a consequence of the 
Parker instability as the dominant mode of fragmentation \citep{Mouschovias09,Koertgen18L,Koertgen19b}. After one 
additional disc rotation the shape of the galaxies is markedly different. The hydrodynamic disc has 
broken up into many almost spherical objects. The magnetised galaxy shows more filamentary structures. 
Here, the magnetic field restricts the motion of the diffuse gas. Hence, gas is not arbitrarily 
absorbed by the formed clouds. We emphasise the presence of spiral features around some 
clouds in both galaxies due to the cloud's bulk rotation. At even later times, the global disc 
morphology looks very similar. This indicates that, at this stage, the magnetic field does not have 
a pronounced impact on the global dynamics of the disc. Filamentary structures are now also seen in 
the hydrodynamic galaxy as a consequence of cloud-cloud interactions.\\
In the top row of Fig.~\ref{fig:glovie_vdisp} we show the divergence of the velocity field at a time close to the end of the simulation for both discs (dark: diverging, bright: converging). While this 
operation on the residual velocity field generates filamentary structures, there still 
appear pronounced differences. For instance, the hydrodynamic galaxy reveals a much 
more converging velocity field throughout the disc. In contrast, the magnetised galaxy 
reveals large patches of diverging gas in between kpc-scale filaments with a 
primarily converging field. In addition, the hydrodynamic disc allows 
for the formation of small scale bundles in the divergence field (see e.g. the pattern near $x=+6\,\mathrm{kpc}$ and $y=+2\,\mathrm{kpc}$), 
whose formation is suppressed in the magnetised galaxy.\\
In the bottom panel of Fig.~\ref{fig:glovie_vdisp}, we show the vorticity. Here, again, filamentary structures can be identified. In contrast to the divergence field, these 
are not as thick, indicating that the vorticity throughout the disc seems to show 
a quite common orientation. In general, regions of pronounced changes in the 
divergence or the vorticity correlate well with gas structures emerging from 
dynamical interactions, such as cloud-cloud collisions. However, both discs do not show
any systematic trends e.g. with respect to the distance from the center of the galaxy.

\subsection{Time evolution of cloud properties}
In this section we focus on the time evolution of dynamical quantities of the found clouds. Before we commence, we describe the 
cloud tracking algorithm.
\subsubsection{Method to track the clouds}
At a given time, a domain of size \mbox{$L_\mathrm{x}\times L_\mathrm{y}\times L_\mathrm{z}=20\times20\times2\,\mathrm{kpc}^3$} 
centered on the galaxy
is 
searched for connected regions. The regions are defined by a minimum density, $n_\mathrm{min}=100\,\mathrm{cm}^{-3}$, which 
is about the average number density of molecular clouds in the Milky Way \citep{Blitz07,Dobbs14}. A grid cell belongs to a certain 
structure if its density is similar (provided some uncertainty) to the one of its nearest neighbours and further has a 
spatial connection to them. The search starts in the cell with the highest density. Once, all objects with 
$n_\mathrm{min,obj}\geq n_\mathrm{min}$ have been found, their current center of mass position and bulk velocity are used to 
follow their evolution in time. At the next timestep -- we use $\Delta t=2\,\mathrm{Myr}$ --, a volume of $V=1\,\mathrm{kpc}^3$ centered on the predicted center of mass is searched 
for clouds and the cloud with its center of mass closest to the predicted one is chosen \citep[see also][]{Tasker09}. We note here that, in the current 
version and as a difference to \citet{Tasker09} and \citet{Jin17}, the cloud tracking algorithm does not search for newly formed clouds and thus traces only the evolution of the 
firstly identified ensemble of clouds.

\subsubsection{Comparing cloud dynamical properties}
\begin{figure*}
    \centering
    \includegraphics[width=0.45\textwidth]{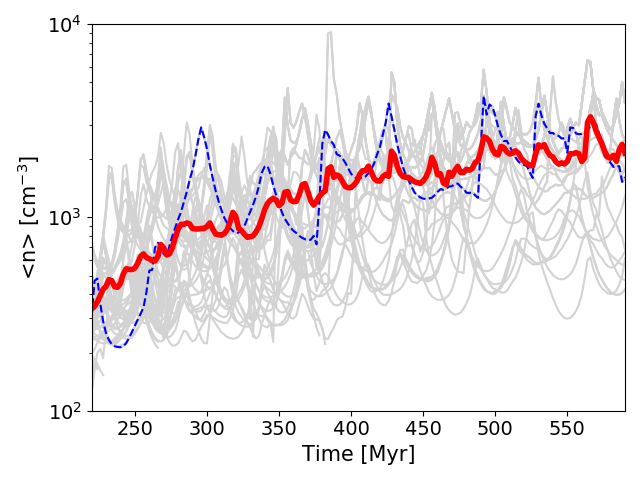} \includegraphics[width=0.45\textwidth]{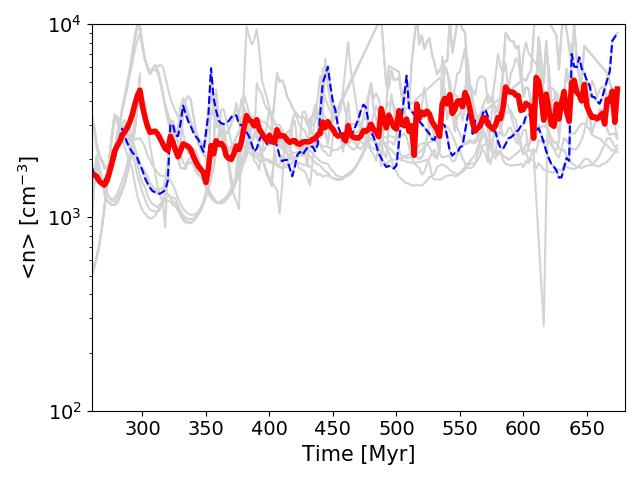}
    \caption{Average number density of the tracked molecular clouds as a function of time. In grey we show the 
    evolution of each individual cloud, while the thick red line denotes the average over all clouds. The blue 
    dashed lined highlights the evolution of a single cloud. Left for the hydrodynamic case. Right for the 
    magnetised case.}
    \label{fig:clprop_dens}
\end{figure*}

\begin{figure*}
    \centering
    \includegraphics[width=0.45\textwidth]{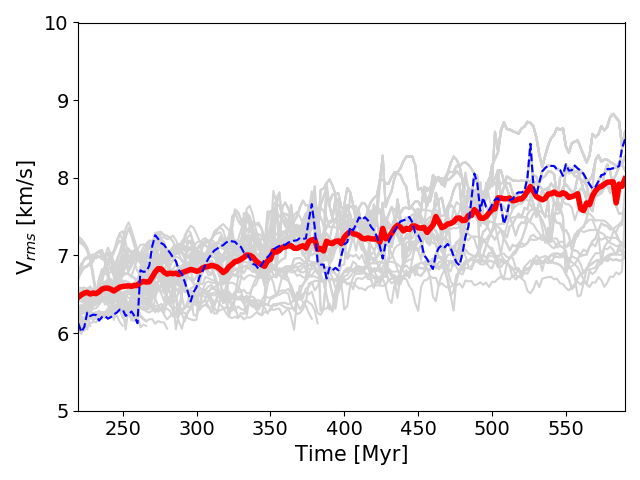} \includegraphics[width=0.45\textwidth]{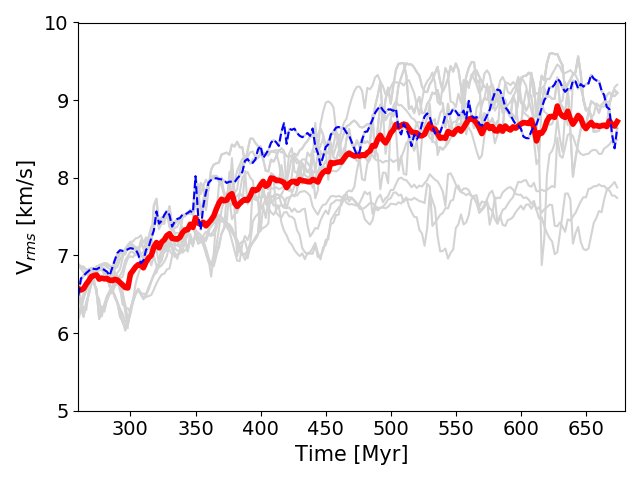}
    \caption{Velocity dispersion of the tracked molecular clouds as a function of time. In grey we show the 
    evolution of each individual cloud, while the thick red line denotes the average over all clouds. The blue dashed lined highlights the evolution of a single cloud. Left for the hydrodynamic case; right for the magnetised case.}
    \label{fig:clprop_mach}
\end{figure*}

If not stated differently, the left panel in the following discussion corresponds to the hydrodynamic disc, while the right 
panel indicates results of the magnetised galaxy. In all figures, the grey lines indicate the evolution of each individual 
cloud, the red thick line is the cloud-mass weighted average, and the blue dashed lines represents the evolution of an 
example cloud\footnote{To be more precise, it is the same cloud in all plots.}.\\
In Fig.~\ref{fig:clprop_dens} we show the time evolution of the average number density of the clouds. The large difference in 
density between the two scenarios is initially due to the higher density of the magnetised galaxy, which is necessary to keep the Toomre parameter at $Q=2$. Apart from this initial difference, the long term evolution of the cloud's average 
density is very similar. There appear, however, subtle variations, i.e. the magnetised clouds reach higher densities of up to 
$\left<n\right>\sim10^4\,\mathrm{cm}^{-3}$, whereas there is only one cloud at a single time in the hydrodynamic disc, which reaches 
such high average densities. The \ita{average} instead is less affected, with a difference of about a factor of two. The 
overall time evolution of the average is more pronounced in the hydrodynamic disc and a plateau is reached after about 
200 Myr of evolution (i.e. one disc rotation). As can be seen from the evolution of the example clouds, both ensembles are 
affected by cloud-cloud interactions with actual collisions resulting in a sudden increase of the average density\footnote{We point out that the impact of the artificial pressure term, which 
results in an adiabatic behaviour in the innermost region of the core, does not affect the 
resulting (average) cloud radii and densities}. Interestingly, 
this effect is seen for almost every cloud. However, collisions/mergers induce only a temporarily limited increase in the 
average density and it takes several interactions for the cloud to become denser on average with time.\\
The time evolution of the velocity dispersion\footnote{Derived from a size-linewidth relation due to 
the limited numerical resolution and as a guideline for the internal dynamics.} is shown 
in Fig.~\ref{fig:clprop_mach}. As expected for unimpeded gravitational collapse, the velocity 
dispersion increases over time. The retrieved velocity dispersion evolves 
very similar in both galaxies, but is shifted to larger values in the MHD disc. The scatter, in 
contrast, is larger in 
the hydrodynamic case, which represents a larger variety of cloud properties.  While it takes about one 
disc rotation to reach saturation in the magnetised 
disc, the hydrodynamic clouds still show a net increase over the course of about one and a half 
disc rotations and have not reached a saturated stage yet. Given the fact that the velocity dispersion is 
calculated from the size-linewidth relation $\sigma_v\sim1.1R^p$, with $p=0.38$ and $R$ being the cloud-size \citep{Larson81}, the saturation indicates a phase in which the clouds 
do not grow anymore. The typical cloud temperature is about $T\sim30\,\mathrm{K}$, which corresponds 
to a sound speed of about $c_s\sim0.4\,\mathrm{km/s}$. Therefore, sonic Mach numbers, as derived from 
Fig.~\ref{fig:clprop_mach}, are in the range of $\mathcal{M}_s\sim15-25$.

\subsubsection{Additional dynamics of the magnetised clouds}
\begin{figure*}
    \centering
    \includegraphics[width=0.45\textwidth]{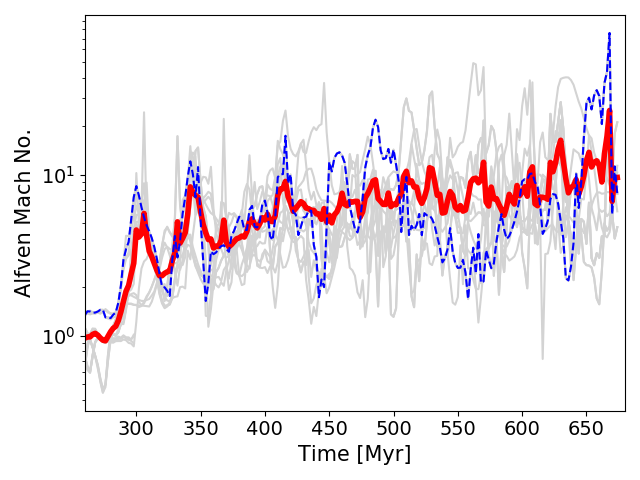} \includegraphics[width=0.45\textwidth]{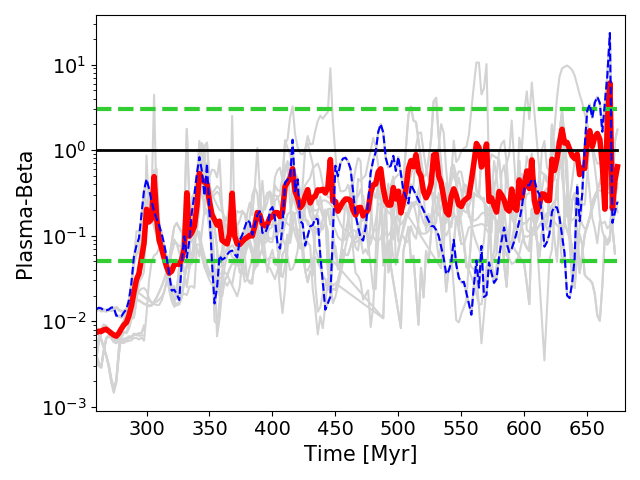}
    \caption{\ita{Left:} Alfv\'{e}n Mach number of the tracked molecular clouds as a function of time. \ita{Right:} Plasma-$\beta$ as a function of time. The 
    line colors and styles have the usual meaning. In the right sub-figure, the 
    black line denotes $\beta=1$ and the green dashed lines indicate upper and lower 
    limits for the magnetisation of solar neighbourhood clouds as constrained in 
    \citet{Kainulainen17}.}
    \label{fig:clprop_malfbeta}
\end{figure*}

Fig.~\ref{fig:clprop_malfbeta} emphasises the time evolution of the cloud's 
Alfv\'{e}n Mach number and plasma-$\beta$. Similar to the sonic Mach number, the 
average Alfv\'{e}n Mach number appears to reach saturation after one disc rotation. 
However, as inferred from the example cloud, individual clouds can show 
dramatic changes in the dynamic interplay of turbulence and magnetic fields. In 
general, all clouds are seen to be super-Alfv\'{e}nic. This is also a 
result of matter accretion along the field lines, which results in a net decrease 
of the Alfv\'{e}n speed when the field is not amplified sufficiently. Please note 
that the identified clouds start out trans- to slightly
super-Alfv\'{e}nic\footnote{We expect the initial conditions to be sub- to 
trans-Alfv\'{e}nic for increasing resolution.} and reach values of 
$\mathcal{M}_A\gtrsim3$ after $\sim40\,\mathrm{Myr}$. The ratio of thermal to magnetic 
pressure of the clouds varies between $\beta\sim0.01-1$ for the majority of the 
clouds, with a rather stable average value of $\left<\beta\right>\sim0.2-0.4$, which
 is also in agreement with the constraints discussed by \citet{Kainulainen17} for 
 a set of solar neighbourhood clouds. Furthermore, the average value also fits the 
 global disc average \citep{Koertgen19b}.
\subsection{The density variance - Mach number relation}
\begin{figure*}
    \centering 
    \includegraphics[width=0.45\textwidth]{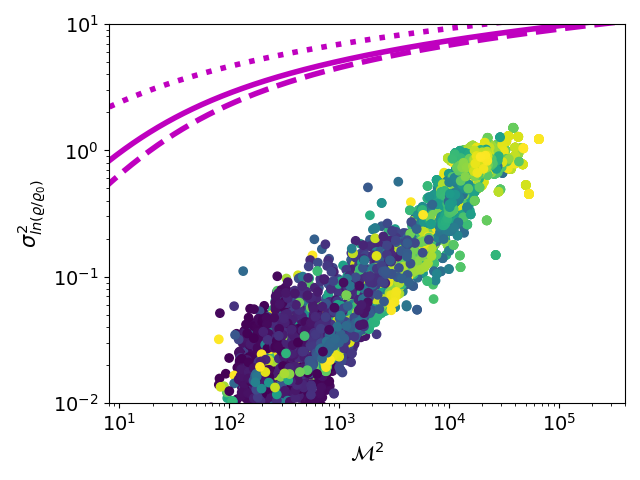}
    \includegraphics[width=0.45\textwidth]{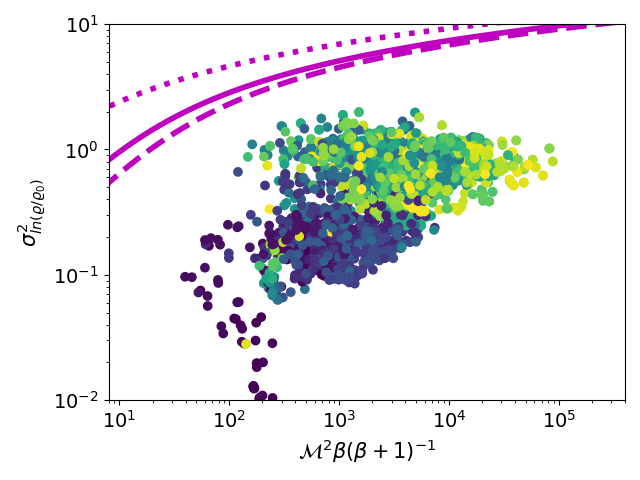}
    \caption{Density variance -- Mach number relation at different times 
    (blue: early, yellow: late). In the right panel, we show the magnetised 
    version, which accounts for the plasma-$\beta$. Although the magnetised 
    disc is closer to the theoretical prediction, no relation between the 
    density variance and the Mach number term arises. In contrast, a clear 
    relation is seen in the hydrodynamic case. However, the non-converged 
    velocity field and the resulting inaccurate Mach numbers yield a 
    relation that is far off the theoretical ones (purple lines. Dotted: $b=1$, solid: $b=0.4$, dashed: $b=1/3$).}
    \label{fig:densvarmach}
\end{figure*}

Having the time evolution of the relevant quantities at hand, one can investigate whether there 
emerges a relation between these and the density-variance.
For the logarithmic density contrast this relation is 
given by \citep[e.g][and references therein]{Federrath12}
\beq
\sigma_{\mathrm{ln}\left(\varrho/\varrho_0\right)}^2=\mathrm{ln}\left(1+b^2\mathcal{M}^2\frac{\beta}{\beta+1}\right).
\label{eq:densvarmach}
\eeq
We stress here that we remove systematic motions in the velocity field of the cloud and correct for 
a gradient in density, which would affect the density variance \citep{Federrath11a}. 
Fig.~\ref{fig:densvarmach} shows the resulting density variance - Mach number relation for the hydrodynamic 
and magnetised discs. We show all ensemble clouds at all times, where the time is colour coded. As is observed, a relation emerges in the hydrodynamic case, which is, however, 
not described by eq.~\ref{eq:densvarmach}. At first, the density variance is too low for the 
estimated Mach numbers and, secondly, the relation appears rather linear than logarithmic. 
Interestingly, in the 
magnetised counterpart, a logarithmic relation is slightly recognised, which indicates that the 
overall relation is changed quite significantly by the magnetic field. The spread 
across the clouds is larger, but the overall density variances come closer to the theoretical 
values. However, for individual snapshots in time, no significant correlation would be observed. 
This is especially clear for the MHD disc, where the majority of data points at late times (yellow) 
show a rather flat distribution, while the hydrodynamic clouds set up a point cloud between 
$\mathcal{M}^2\sim10^4-10^5$ without any clear trend. In any case, in both discs the derived driving 
parameters would be far too low. This has 
recently been discussed in \citet{Jin17}. The authors showed that the derived $b$ is only a lower 
limit and it was seen to increase with increasing numerical resolution. We stress here that also no 
correct relation emerges, when using the velocity dispersion from the size-linewidth relation. 
\subsection{The turbulence driving parameter derived from the compressive ratio}
\begin{figure*}
    \centering
    \includegraphics[width=0.45\textwidth]{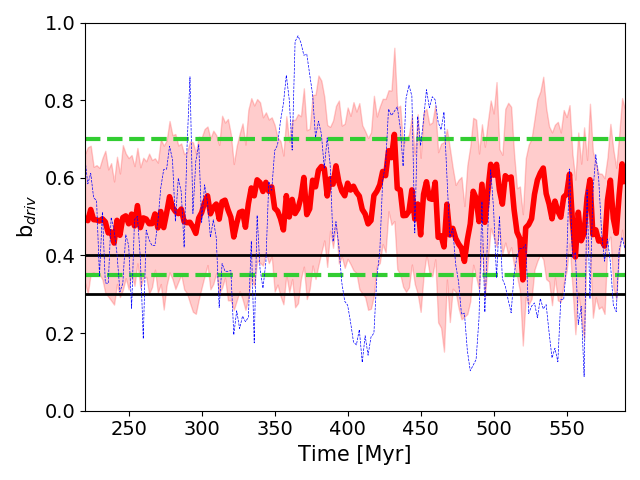} \includegraphics[width=0.45\textwidth]{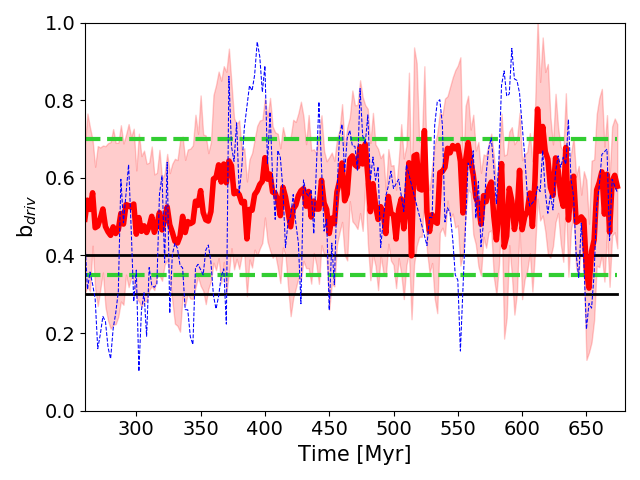}
    \caption{Turbulence driving parameter $b$ as derived via the compression ratio $\chi$ as a function of time for 
    the ensemble of clouds identified in the hydrodynamic (left) and MHD 
    (right) disc. Although strong variations per cloud from purely solenoidal 
    to largely compressive are seen, the mass-weighted average driving 
    parameter is about $b\sim0.5$, indicating slightly compressive driving. In blue 
    we show the evolution of $b$ for an exemplary cloud. There are times with 
    little variation and events, where $b$ changes dramatically within only a few Myr. The stronger 
    fluctuations are due to cloud collisions and tidal interactions.
     The horizontal black lines denote natural ($b=0.4$) and purely solenoidal 
    driving ($b=1/3$), while the green dashed lines highlight the possible upper/lower limit for 
    solar neighbourhood clouds discussed in \citet[][i.e. $b_l=0.35$ and $b_u=0.7$]{Kainulainen17}.}
    \label{fig:bdriv_time}
\end{figure*}

\begin{figure*}
    \centering
    \includegraphics[width=0.45\textwidth,angle=-90]{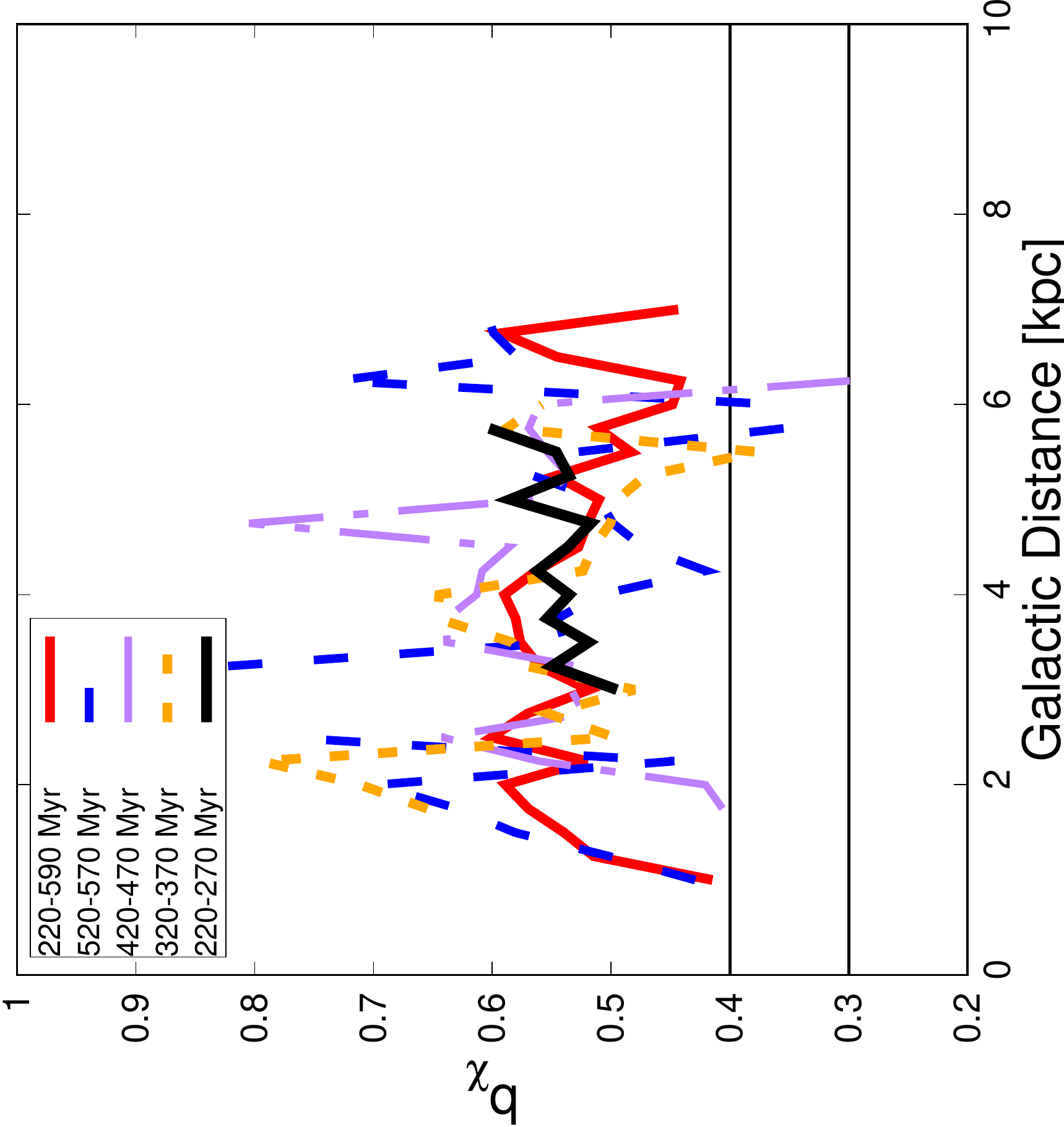} \includegraphics[width=0.45\textwidth,angle=-90]{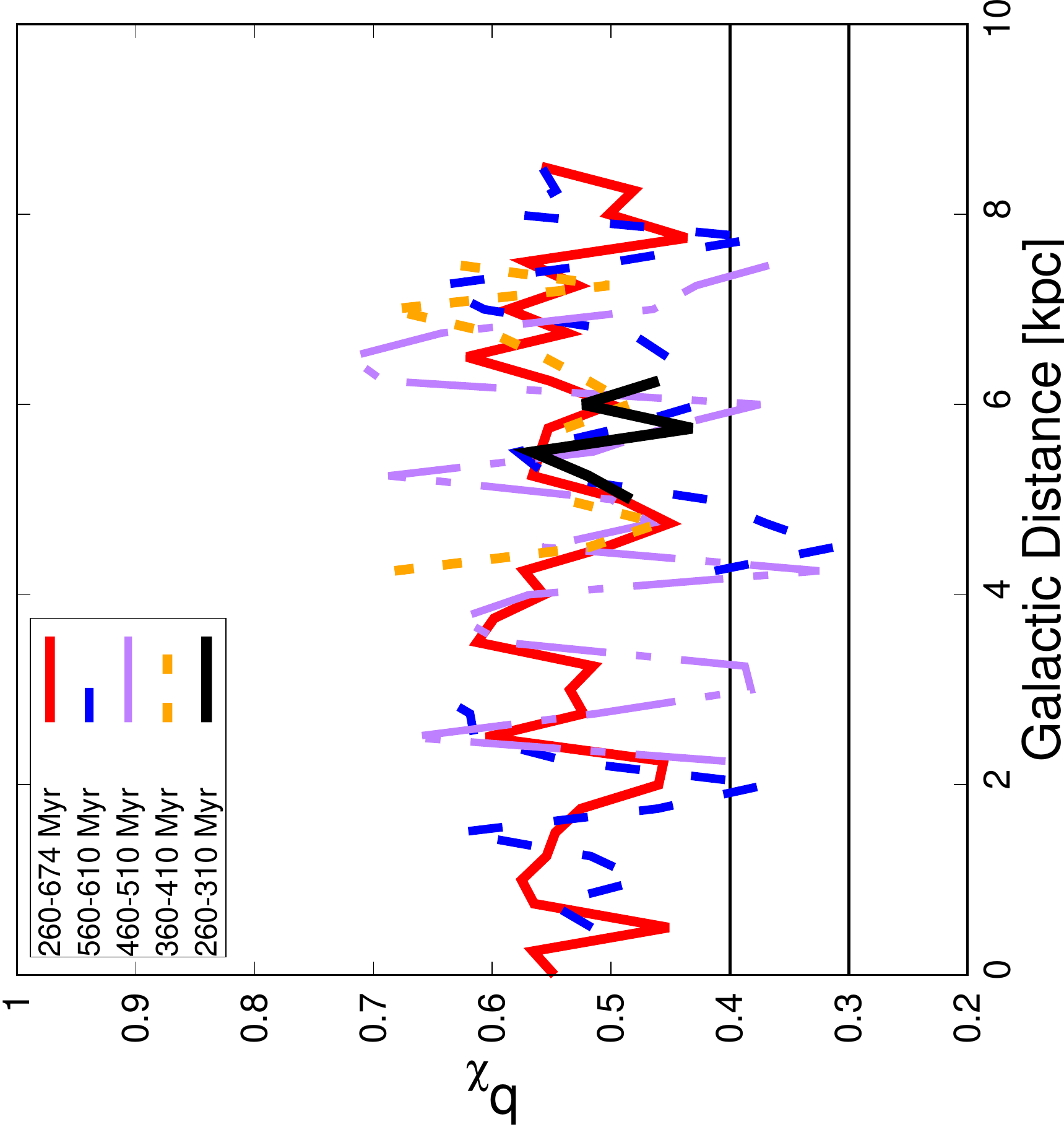}
    \caption{Turbulence driving parameter $b$ as derived via the compressive ratio $\chi$ as a function of 
    distance from the galactic center at various times and averaged over an 
    interval of $\Delta t_\mathrm{av}=50\,\mathrm{Myr}$. Two facts are 
    observed: 1) The driving parameter fluctuates less in the 
    magnetised case and 2) it is generally slightly smaller than the corresponding 
    hydrodynamic case. There is also only small variation in $b$ over 
    the course of a disc rotation. Note that the magnetised clouds are initially 
    less compressive.}
    \label{fig:bdriv_rad}
\end{figure*}

Since Fig.~\ref{fig:densvarmach} indicates the lack of a relation between the density 
variance and Mach number as provided by analytical models of turbulent fragmentation \citep{Molina12}, we refrain from using this relation. We show in the 
appendix below that the density PDFs are dominated by a power-law tail. Hence, the application of the log-normal 
density variance should be taken with caution. However, using instead the density variance for the 
linear density contrast, $\varrho/\varrho_0$,
\beq
\sigma_{\varrho/\varrho_0}^2=b^2\mathcal{M}^2\frac{\beta}{\beta+1}
\label{eq:denslinear}
\eeq
gives 
$b$-values, which are still too low. We re-iterate that the reason for this discrepancy is the lack of numerical resolution in our simulations, as previously found in \citet{Jin17}. We thus concentrate in the 
following on the driving parameter as derived from the compressive ratio 
$\chi=\left<v_\mathrm{comp}^2\right>/\left<v_\mathrm{sol}^2\right>$, where 
$v_\mathrm{comp}$ and $v_\mathrm{sol}$ are the compressive and solenoidal 
components of the velocity field. These can be retrieved via a Helmholtz 
decomposition in Fourier space. It can then be shown \citep{Pan16} 
that the compressive ratio and the turbulence driving parameter are related via
\beq
b_\chi=\sqrt{\frac{\chi}{1+\chi}}.
\eeq
We point out that this relation provides only an approximation to the true value of 
$b$, which can be retrieved from equations \ref{eq:densvarmach} and \ref{eq:denslinear}.
We further note here that we remove any systematic motion in the velocity field, i.e. 
bulk motions in spherical shells around the center of mass. Thus, 
the compressive ratio in this study is comparable to the turbulent compressive 
ratio given in \citet[][see also \citep{Jin17,Mandal20,Menon20}]{Pan16}.\\
The resulting time evolution of the cloud-mass weighted average driving 
parameter, as well as of its standard deviation, is provided in Fig.~\ref{fig:bdriv_time}. Surprisingly, there is almost no pronounced difference 
between the hydrodynamic (left) and magnetised clouds (right). The magnetised 
driving parameter appears to be slightly higher, but is still around 
$b_\chi\sim0.5$\footnote{Note that we use $b_\chi$ and $b_\mathrm{driv}$ interchangeably in the figures.}.  A closer look at the early evolution reveals that the 
magnetised parameter starts fluctuating earlier, after $\sim100\,\mathrm{Myr}$, 
while the hydrodynamic driving parameter keeps its initial value for over 
half a rotation of the galaxy (again estimated at $R=8\,\mathrm{kpc}$). Over the 
course of the evolution, the average $b_\chi$ starts to fluctuate quite strongly, 
thereby taking values from fully solenoidal to strongly compressive. However, 
despite the large fluctuations, both average driving parameters are within the 
bounds estimated by \citet[][green dashed lines]{Kainulainen17}. Given the fact that we only include gravity here, the tendency towards compressive driving is not surprising. Please note that, as summarised in \citet{Federrath17}, stellar feedback 
is also likely to drive compressive turbulent motions.\\
The thin blue 
line underlines the time evolution of the driving parameter for an individual 
cloud. Here, large fluctuations and deviations from the average are evident. For example, the hydrodynamic 
example cloud shows a stage near $t\sim370\,\mathrm{Myr}$, where its turbulence 
would be described as fully compressive. In contrast, not even 30\,Myr later, 
around $t\sim400\,\mathrm{Myr}$, its driving parameter has dropped to 
$b_\chi\sim0.2$, which indicates purely solenoidal driving. Similar short-period 
variations are observed for the example cloud in the magnetised disc. Both 
example clouds furthermore show a long-periodic modulation, with the magnetised 
modulation being less clear. We point out that not all clouds show such clear modulation patterns.
However, the short- and long-period features can be 
interpreted as local and global (i.e. galactic-scale) processes affecting the velocity field in the clouds.\\
Fig.~\ref{fig:bdriv_rad} analyses the behaviour of the turbulence driving mode as a function of distance to the 
galactic center. The various lines denote time averages over a period of 50\,Myr or over the full tracking period. 
Although the spatial distribution of clouds in the hydrodynamic disc is narrower, there is no difference to the 
magnetised ensemble of clouds. In addition, we do not find any systematic trends with position in the galaxy, 
although the large fluctuations hint towards small temporal variations, which could indicate a trend at the given 
time. Contrary to the missing radial trend, a few small differences can be recognised. Initially, the clouds in 
the magnetised disc show a slightly lower driving parameter $0.4<b\lesssim0.5$, whereas the hydrodynamic 
clouds reveal $b\sim0.55$. The fluctuations in $b_\chi$ are larger in the hydrodynamic ensemble, which was also 
recognised in Fig.~\ref{fig:bdriv_time}. This implies that the magnetic field tends to (slightly) decrease the 
overall variation in $b_\chi$.
\subsubsection{What causes the changes in the driving parameter?}
\begin{figure*}
    \centering
    \includegraphics[width=0.5\textwidth,angle=-90]{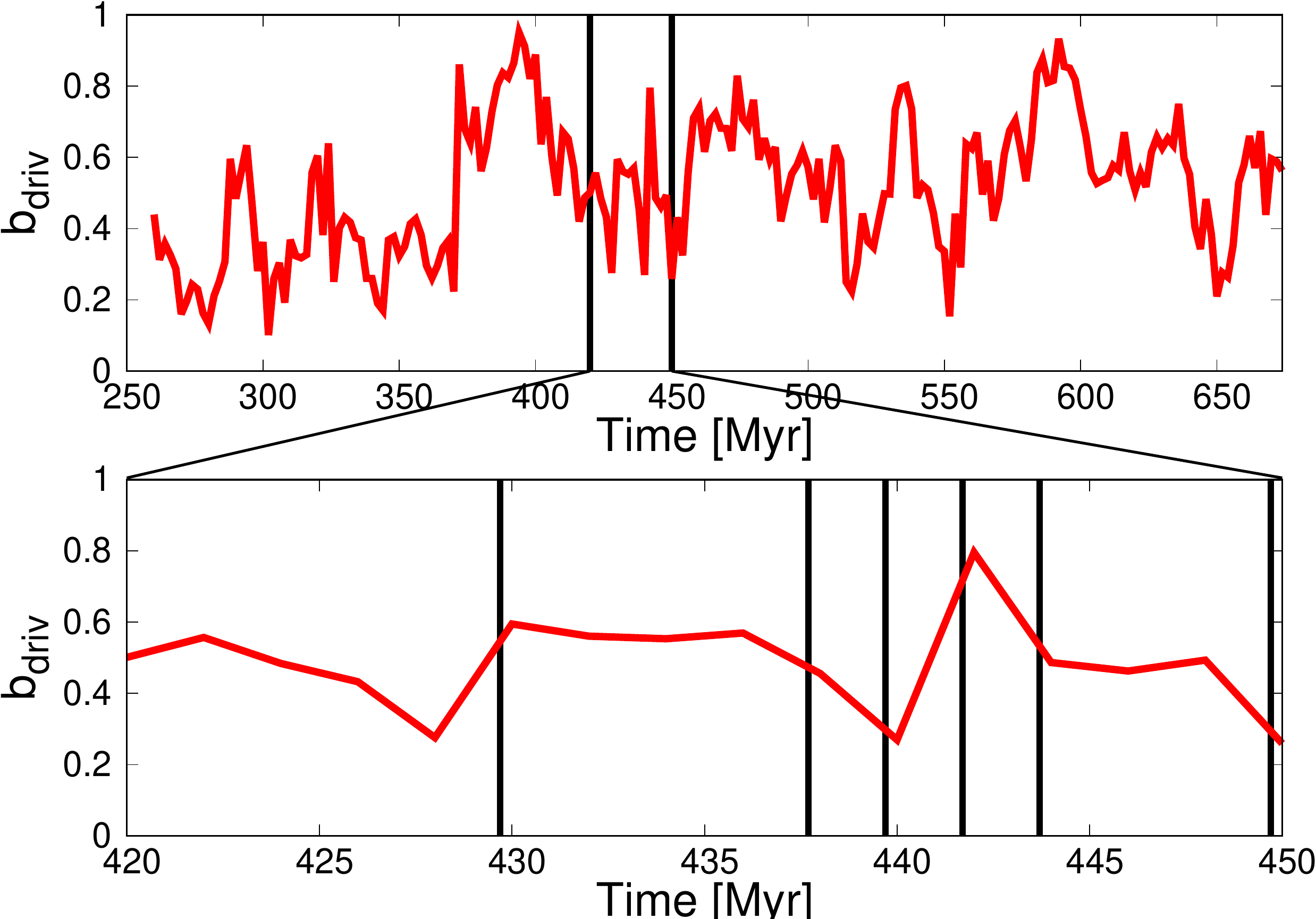}
    \begin{tabular}{lll}
    \includegraphics[width=0.33\textwidth]{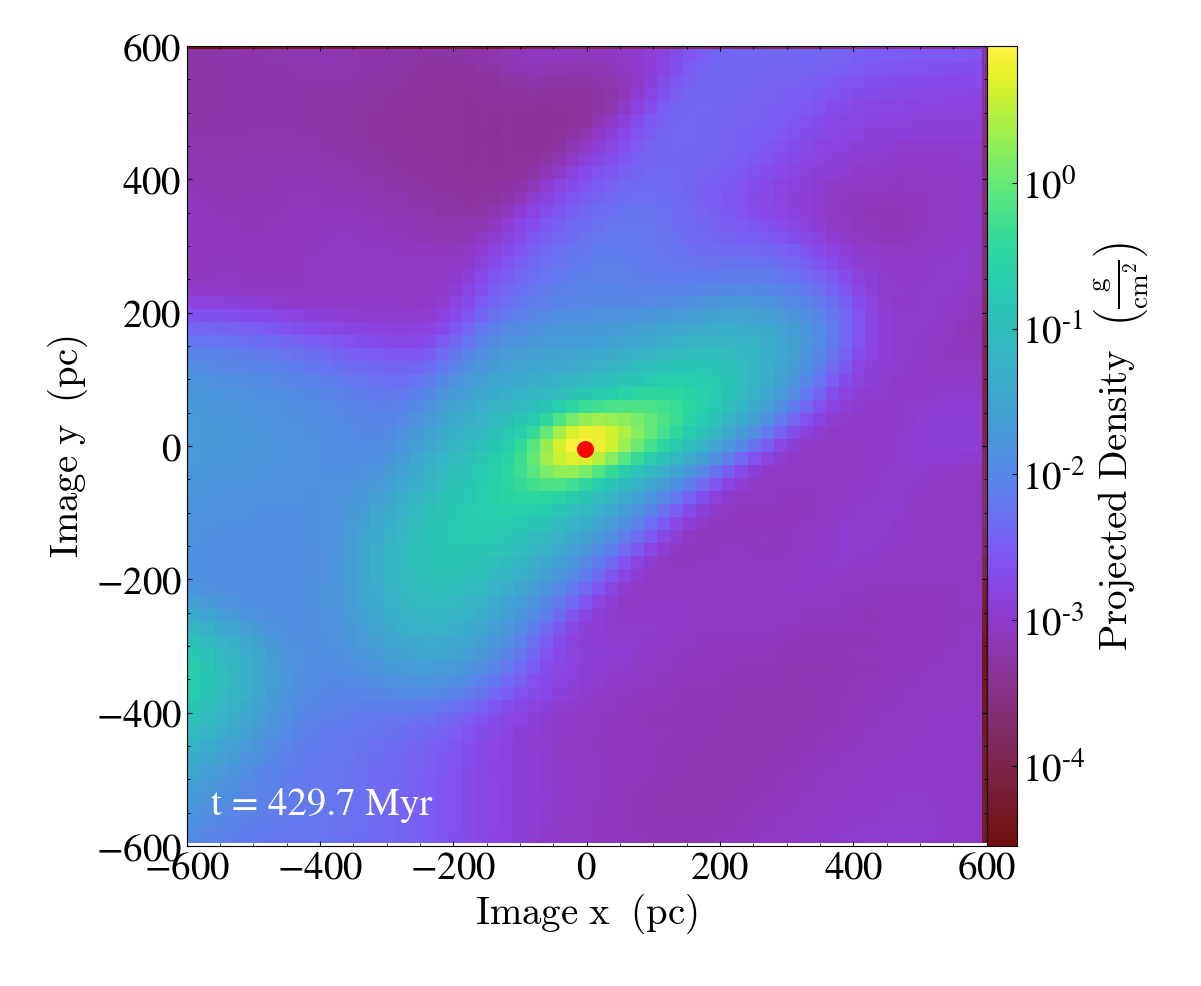}&\includegraphics[width=0.33\textwidth]{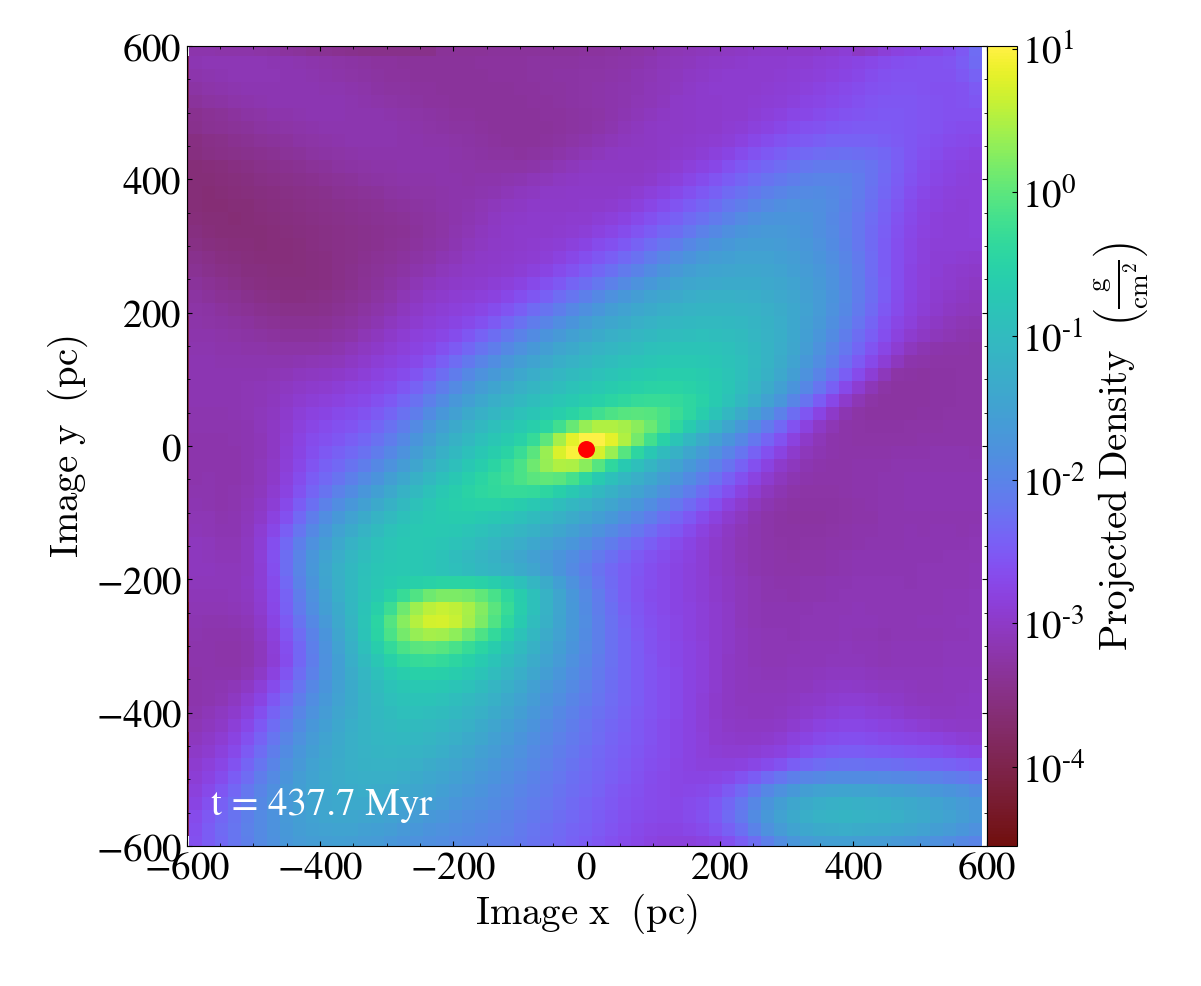}&\includegraphics[width=0.33\textwidth]{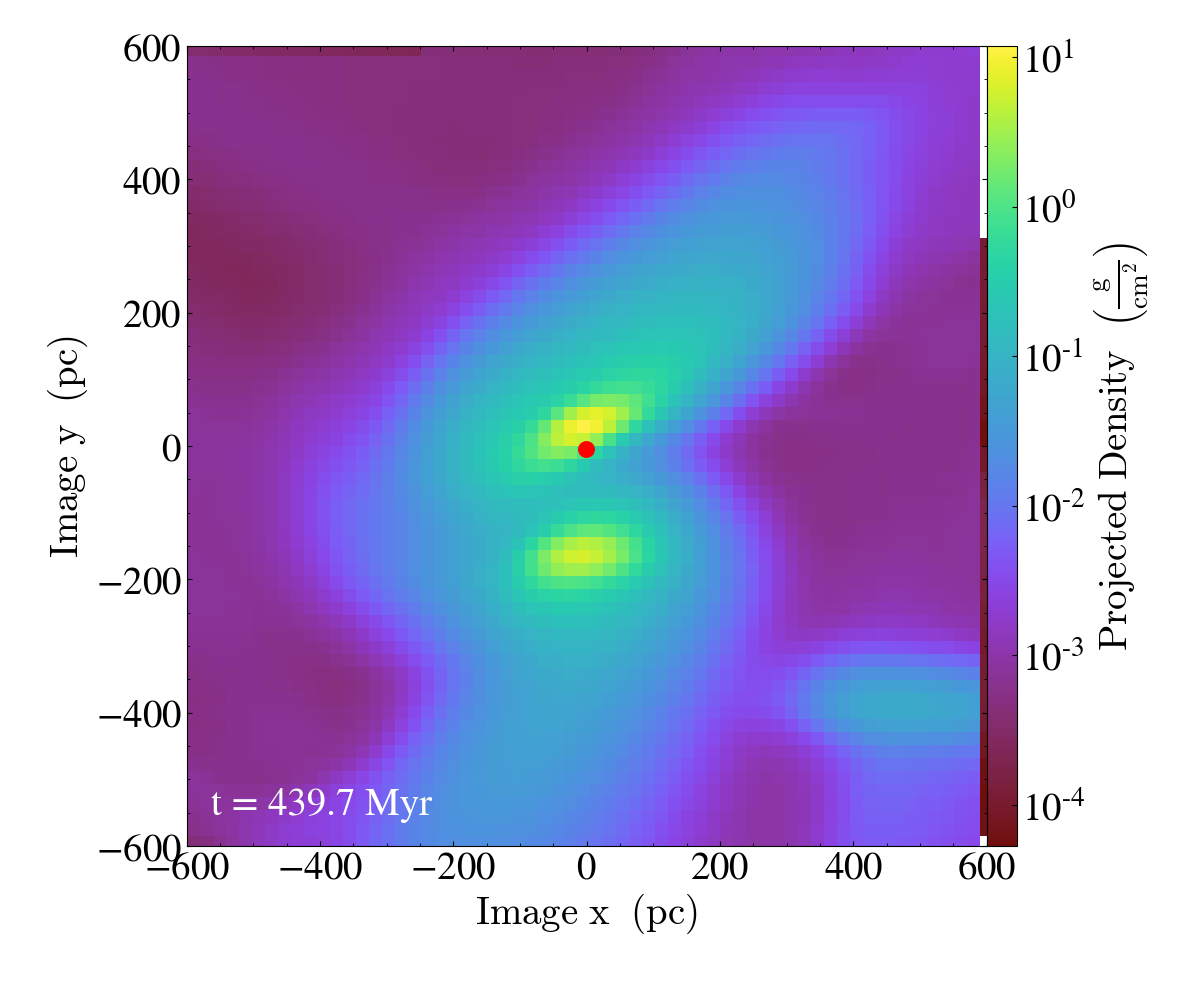}\\
    \includegraphics[width=0.33\textwidth]{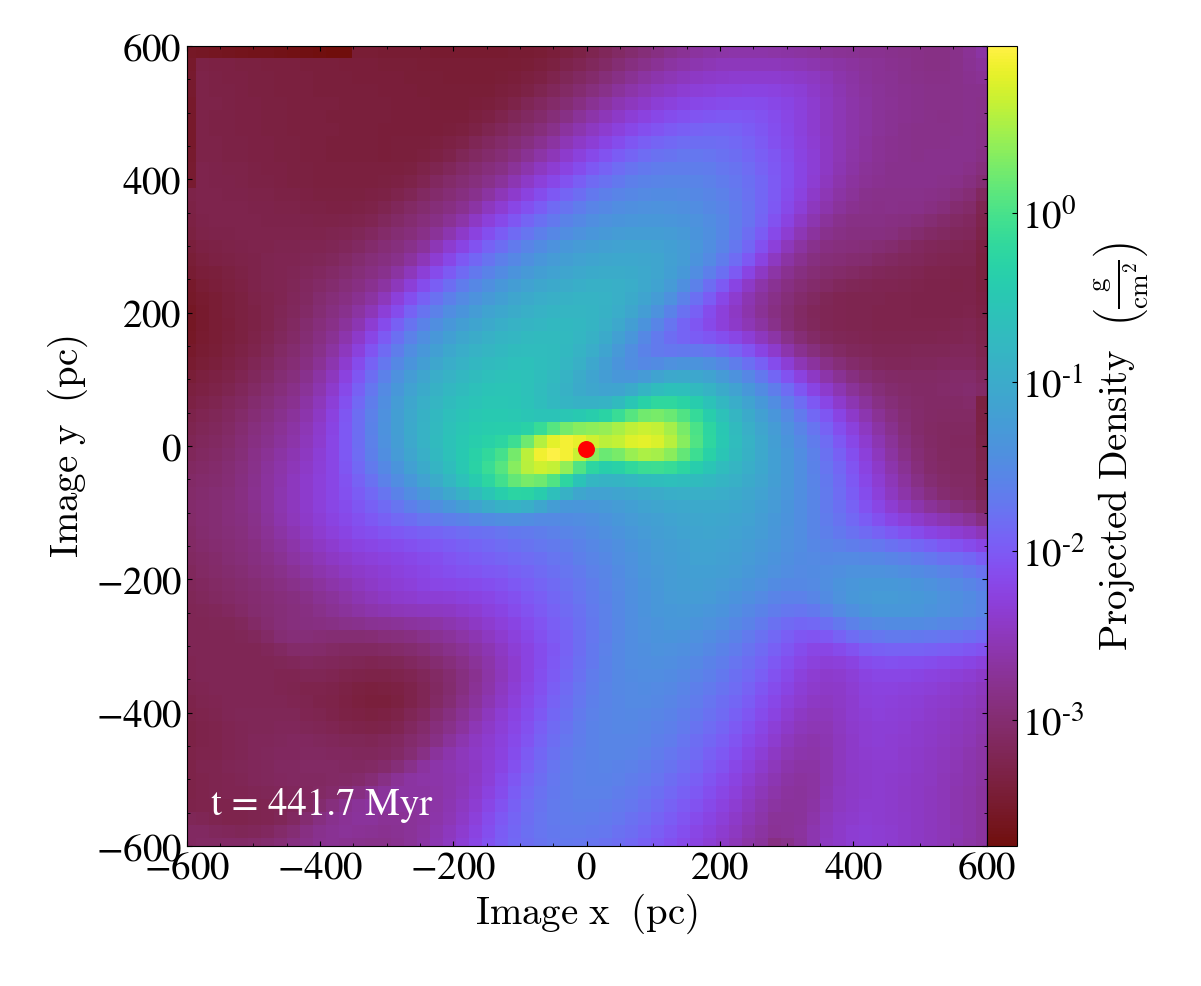}&\includegraphics[width=0.33\textwidth]{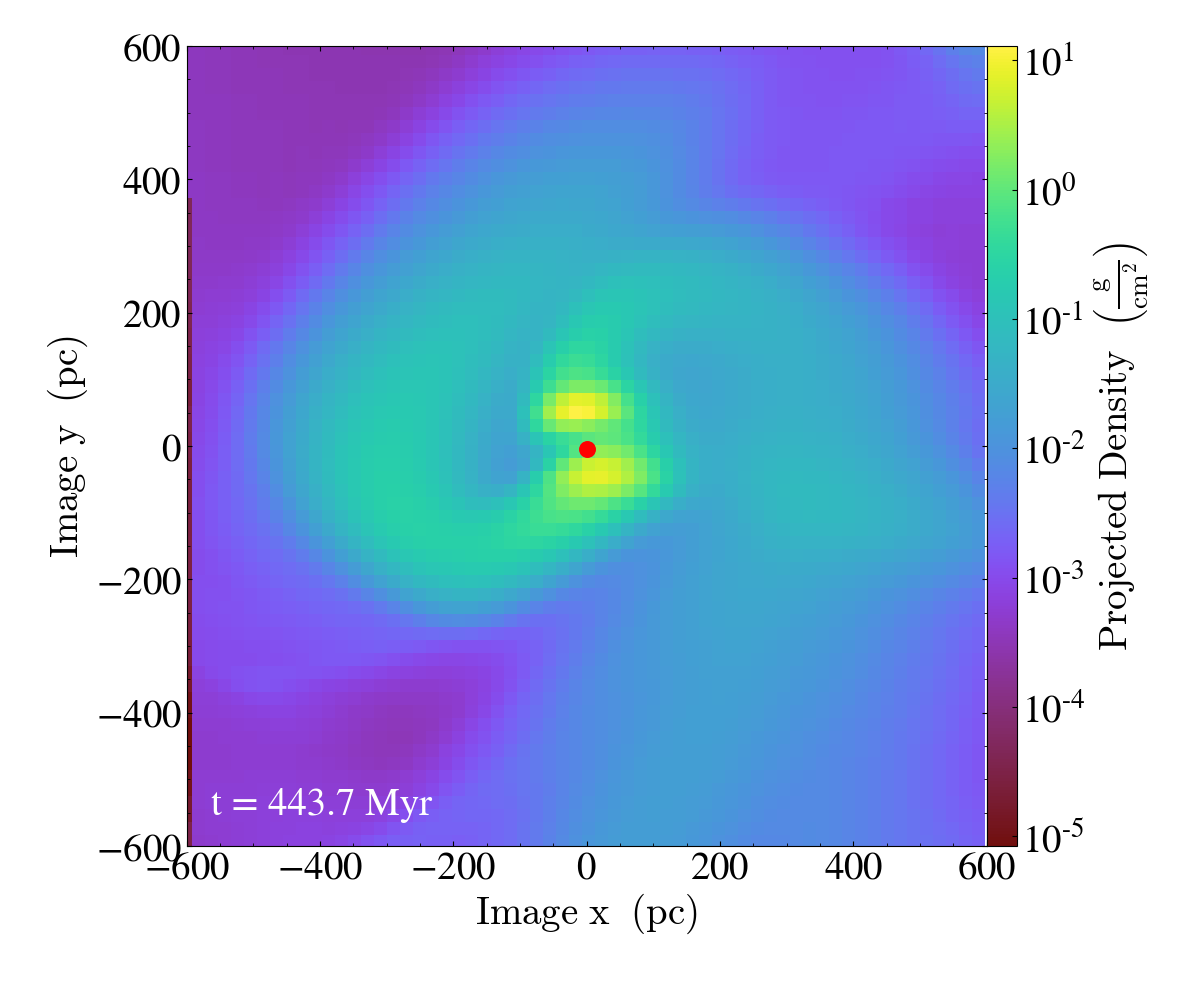}&\includegraphics[width=0.33\textwidth]{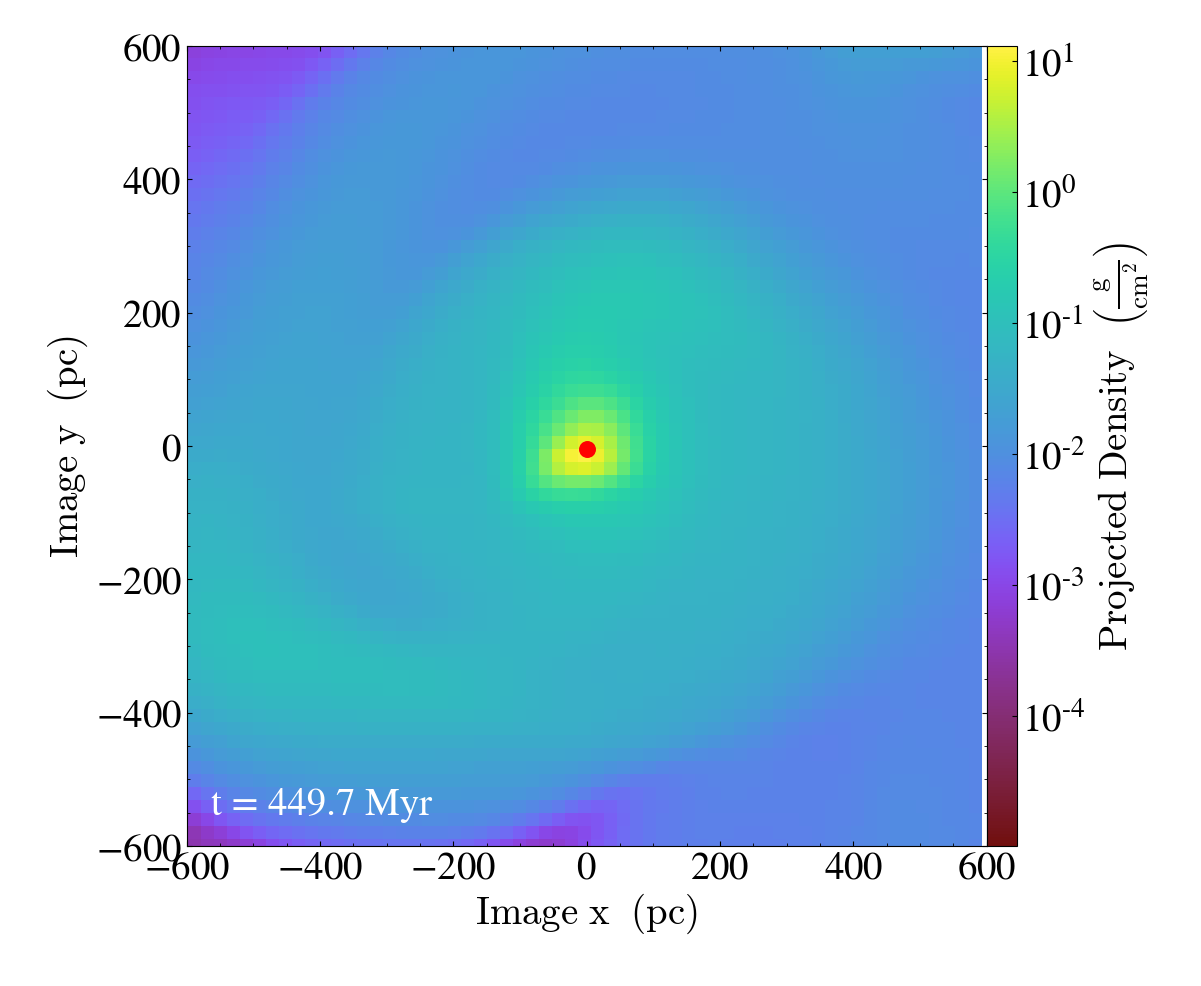}
    \end{tabular}
    \caption{\ita{Top two panels:} Time evolution of the turbulence driving parameter 
    for the example cloud, where a specific time period is highlighted. 
    \ita{Bottom panels:} Surface density maps of a $1.2\times1.2\,\mathrm{kpc}^2$ area centered on 
    the cloud's center of mass (indicated by the red dot). Each panel is highlighted 
    by a vertical line in the time evolution plot (sequence: top to bottom and left to 
    right). The sequence of maps shows a cloud merger and emphasise that this 
    process initiates at first a \ita{decrease} in the driving parameter due to 
    tidal forces inducing shear. The first merger process is highly compressive with 
    $b\sim0.8$. Note the inspiral-phase after the closest approach, which results in 
    strong shearing within the cloud and subsequently in low values of $b$.}
    \label{fig:bdriv_height_time}
\end{figure*}

Keeping in mind the absence of stellar feedback, the lack of a correlation between the driving parameter and the position of the cloud in the galaxy seems to be in 
contradiction with recent estimates of the turbulence driving mode in Galactic molecular clouds, which show mixed to 
mildly compressive turbulence in the solar 
neighbourhood \citep{Ginsburg13,Kainulainen17} and solenoidal forcing near the Galactic center \citep{Federrath16d}. 
Hence, there must be other mechanisms that induce the derived changes in the forcing over time. 
In Fig.~\ref{fig:bdriv_height_time} we show the time evolution of $b_\chi$ for the example cloud. We extract a 
certain interval of about 30\,Myr and present surface density maps centered on the cloud's center of mass. The 
sequence of maps clearly shows that the cloud undergoes a collision/merger with a different, slightly less dense 
object. The corresponding times shown in the surface density maps are highlighted as vertical lines in the 
time evolution of $b_\chi$. The driving mode is initially compressive with $b_\chi\sim0.6$. Interestingly, the 
appearance of the collider reduces the forcing parameter to $b_\chi\sim0.2$. This implies that tidal forces 
induce shear across the cloud, given that the net tidal field is disruptive \citep[see e.g. discussion in][]{Jog13}. The later stages, which resemble an in-spiral phase and the actual collision raise 
the value to $b_\chi\sim0.8$. This is what should be expected from a collision. After the collision event, however, 
the driving parameter decreases again to solenoidal values. This is, because the collision was not head-on and thus 
increased the shear within the cloud. We point out that this change of $b_\chi$ from $\sim0.3$ to $\sim0.8$ and back 
to solenoidal happens on timescales of less than 10\,Myr \citep[in agreement with the findings by][]{Koertgen17b}. Extrapolating this to all clouds, we conclude that the nature
of the collision event is vital for
the short-term temporal variation of the driving parameter.

\subsubsection{Density dependence of the driving parameter}
\begin{figure*}
    \centering
    \includegraphics[width=0.45\textwidth]{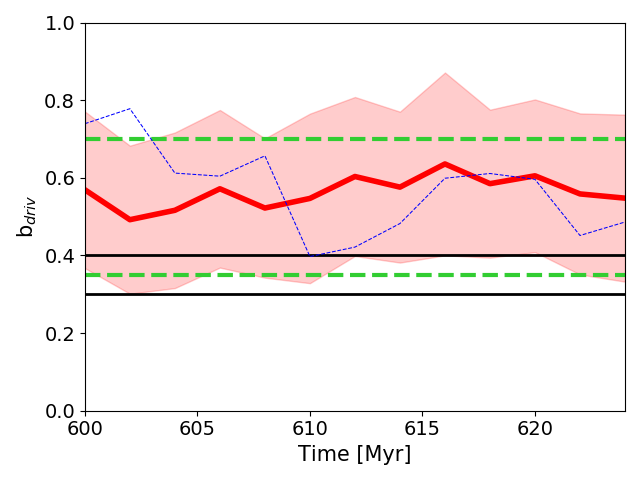}
    \includegraphics[width=0.45\textwidth]{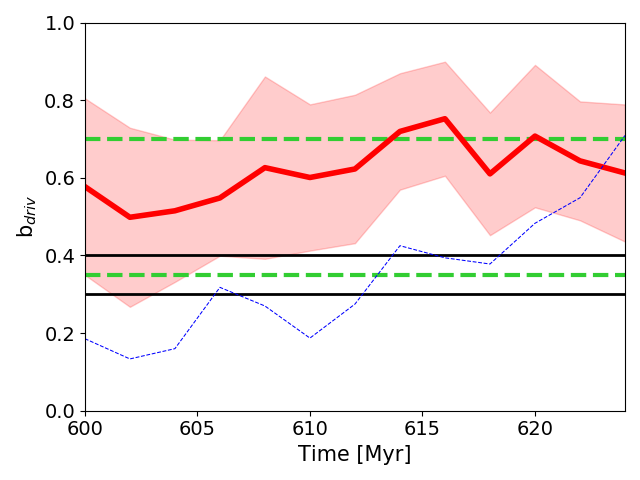}
    \caption{Dependence of the driving parameter $b$ on changing the density threshold 
    for defining clouds in the simulated galaxies. Evolution of identified clouds at 
    late times in the magnetised disc. The left panel shows clouds with a minimum density of 
    $n_\mathrm{min}=100\,\mathrm{cm}^{-3}$, while the right shows clouds with $n_\mathrm{min}=10^3\,\mathrm{cm}^{-3}$. As expected, the clouds with the higher threshold density 
    show a slightly more compressive driving parameter, since these are more gravitationally 
    bound/contracting. This is in agreement with \citet{Orkisz17}, who find a smaller solenoidal fraction on smaller scales. However, the time averages differ only little with $b_\mathrm{cloud}\pm\Delta b_\mathrm{cloud}=0.56\pm0.04$ and $b_\mathrm{dense}\pm\Delta b_\mathrm{dense}=0.62\pm0.07$.}
    \label{fig:bdriv_density}
\end{figure*}

Finally, we briefly study the influence of the choice of the lower threshold density for the identification of 
overdense objects. Fig.~\ref{fig:bdriv_density} shows the time evolution of the average driving parameter of 
about 16 clouds. The left panel shows the standard threshold density of $n=100\,\mathrm{cm}^{-3}$. In the right 
panel we have increased the lower threshold by a factor of ten, thus corresponding to molecular clump or core 
densities. As expected, the driving mode becomes more compressive for the higher threshold density since this 
material is more tightly bound by gravity, although the increase can temporarily be rather 
small \citep[see also][]{Orkisz17}. The average 
driving parameter over this period is $b_\mathrm{cloud}\pm\Delta b_\mathrm{cloud}=0.56\pm0.04$ for the 
fiducial threshold and $b_\mathrm{dense}\pm\Delta b_\mathrm{dense}=0.62\pm0.07$ in case of the increased 
lower density bound. Please note the 
time evolution of the driving parameter for the example clouds. These nicely show that the driving mode can change 
significantly over the course of 25\,Myr and also does not represent the ensemble well.

\section{Summary and conclusions}\label{sect:summary}
We present a study on the time evolution of an ensemble of clouds formed in disc galaxies. We focus on a 
hydrodynamic disc and a strongly magnetised disc with an initial plasma-$\beta=0.25$. The clouds are 
identified shortly after the discs have fragmented on large scales either due to the classical Toomre or the 
Parker instability. We find that the dynamical quantities, relevant for the turbulent properties of the clouds, 
do not vary significantly when including a magnetic field. In a next step, we determined whether 
there arises a relation between the density variance and the turbulent sonic Mach number. Both 
galaxies show some kind of a relation. This, however, does not match the theoretical expectation 
for isothermal, turbulent gas, primarily because of the lack of numerical resolution in our simulations. As a consequence, we do not determine the driving parameter from 
the density variance - Mach number relation, but rather via the compressive ratio. Our main findings 
concerning the properties and evolution of the driving parameter obtained this way are:
\begin{itemize}
    \item The derived driving parameter varies between fully solenoidal ($b_\chi\sim1/3$) 
    and entirely compressive ($b_\chi\sim1$) driving.
    \item The \ita{average} driving parameter, $b_\chi$, does not significantly vary over the evolution of 
    about $\sim400$\,Myr.
    \item In contrast, for individual clouds, we find large fluctuations as well as times of 
    little to almost no variation.
    \item The largest fluctuations of $b_\chi$ can be associated with external distortions such 
    as cloud-cloud mergers.
    \item Collisions between clouds induce a rapid change of $b_\chi$ within only a few Myr.
    \item The tidal forces exerted onto each cloud by its environment can reduce the driving parameter as they increase the shear.
    \item In our current framework, there appear no variations across the disc.
    \item Taking the magnetic field into account slightly reduces the spread/the uncertainty in $b_\chi$.
\end{itemize}
We conclude that the merger history and/or the environment play a significant role in 
shaping the turbulent velocity field of molecular clouds in galaxies.
\section*{Acknowledgement}
BK thanks the anonymous referee for her/his comments, which helped to clarify the findings of this study.
BK acknowledges Paris-Saclay University's Institut Pascal program "The Self-Organized Star Formation Process" and the Interstellar Institute for hosting discussions that nourished the development of the ideas behind this work. 
BK further acknowledges discussions with Robi Banerjee, Urs Sch\"afer, Wolfram Schmidt, Simon Selg and Pranjal Trivedi (in 
alphabetical order) and funding via the Australia-Germany Joint Research Cooperation Scheme (UA-DAAD) and from 
DFG grant BA3706/15-1. The simulations were run on HLRN-III under project grant hhp00043.
The \textsc{flash} code was in part developed by the DOE-supported ASC/Alliance Center for Astrophysical Thermonuclear Flashes at the University of Chicago. Figs.~\ref{fig:glovie_cdens} and \ref{fig:bdriv_height_time} were generated with \ita{GUFY}, a 
graphical user interface for the analysis of \textsc{flash} data with yt, developed by F.~Balzer (University of 
Hamburg).

\begin{appendix}
\section{Density Probability Distribution Function}
\begin{figure}
    \centering
    \includegraphics[width=0.45\textwidth,angle=-90]{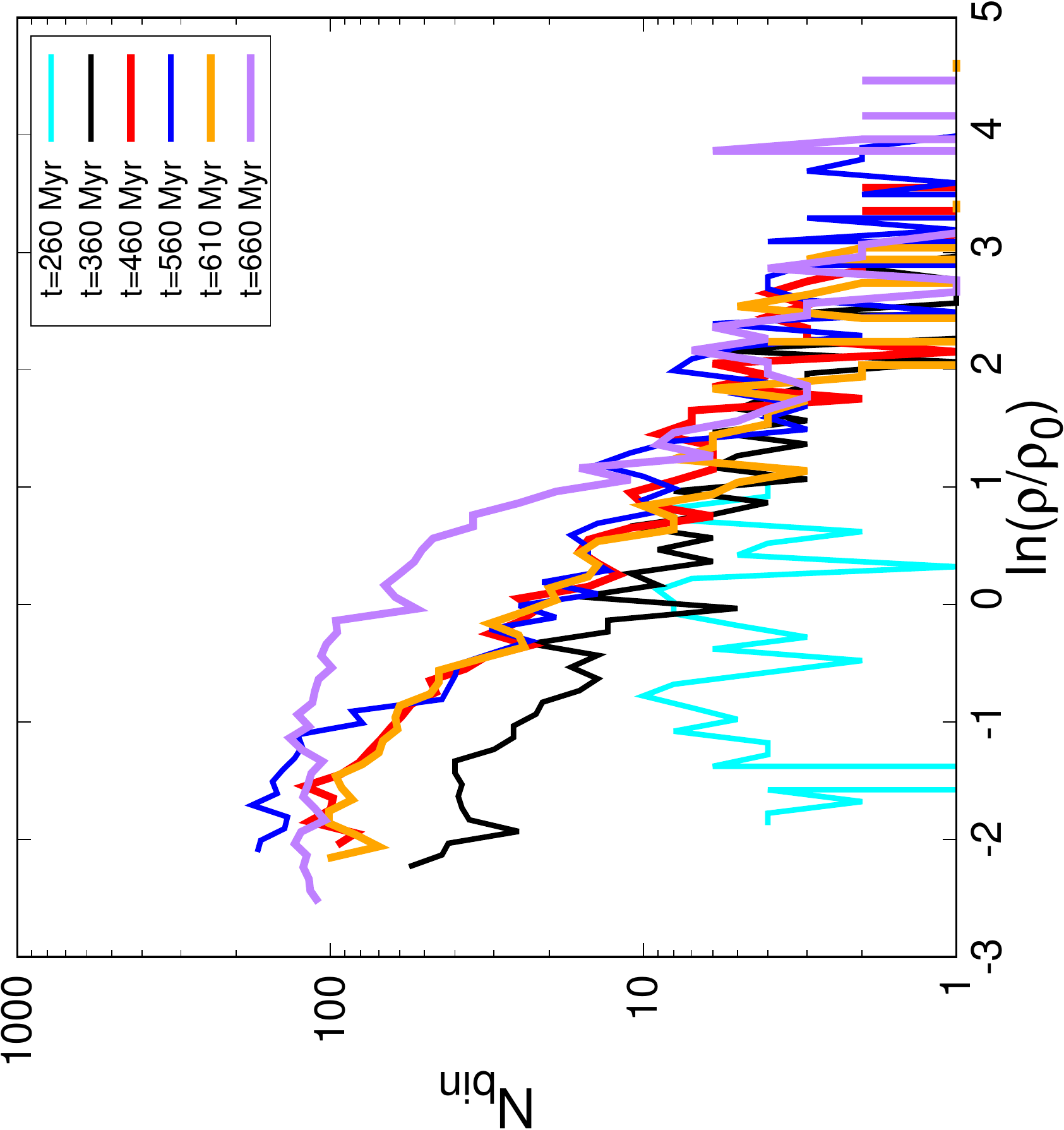}
    \caption{Probability distribution function (PDF) of the logarithmic density contrast 
    for the example cloud at different times. The PDF is initially flat, because the 
    cloud has just condensed out of the diffuse gas and has not developed any 
    substructure, yet. With time, the PDF is transformed into a mixture of a 
    log-normal and a power-law part. However, the power-law part dominates the PDF at 
    most times, which is attributed to the rather coarse numerical resolution, which 
    does not allow the cloud to fragment any further.}
    \label{fig:app_densitypdf}
\end{figure}

A way to study the dynamics of molecular clouds is by analysing their (column-) density distribution. 
The results are shown in Fig.~\ref{fig:app_densitypdf} for different times. The PDF is almost flat 
at the earliest time, when the cloud has just formed via compression of gas flows due to the 
Parker instability. Over time, the PDF develops a power-law shape. At some times, a clear 
log-normal part is observed. In any case, it is clear that the 
PDF is more likely to be a power-law than a combination of lognormal and power-law, probably due to the 
lack of resolution. Hence, 
deriving the turbulence driving parameter from the logarithmic density variance - Mach number 
relation should be taken with caution.

\section{Resolution study}
\begin{figure*}
    \centering
    \includegraphics[width=0.45\textwidth,angle=-90]{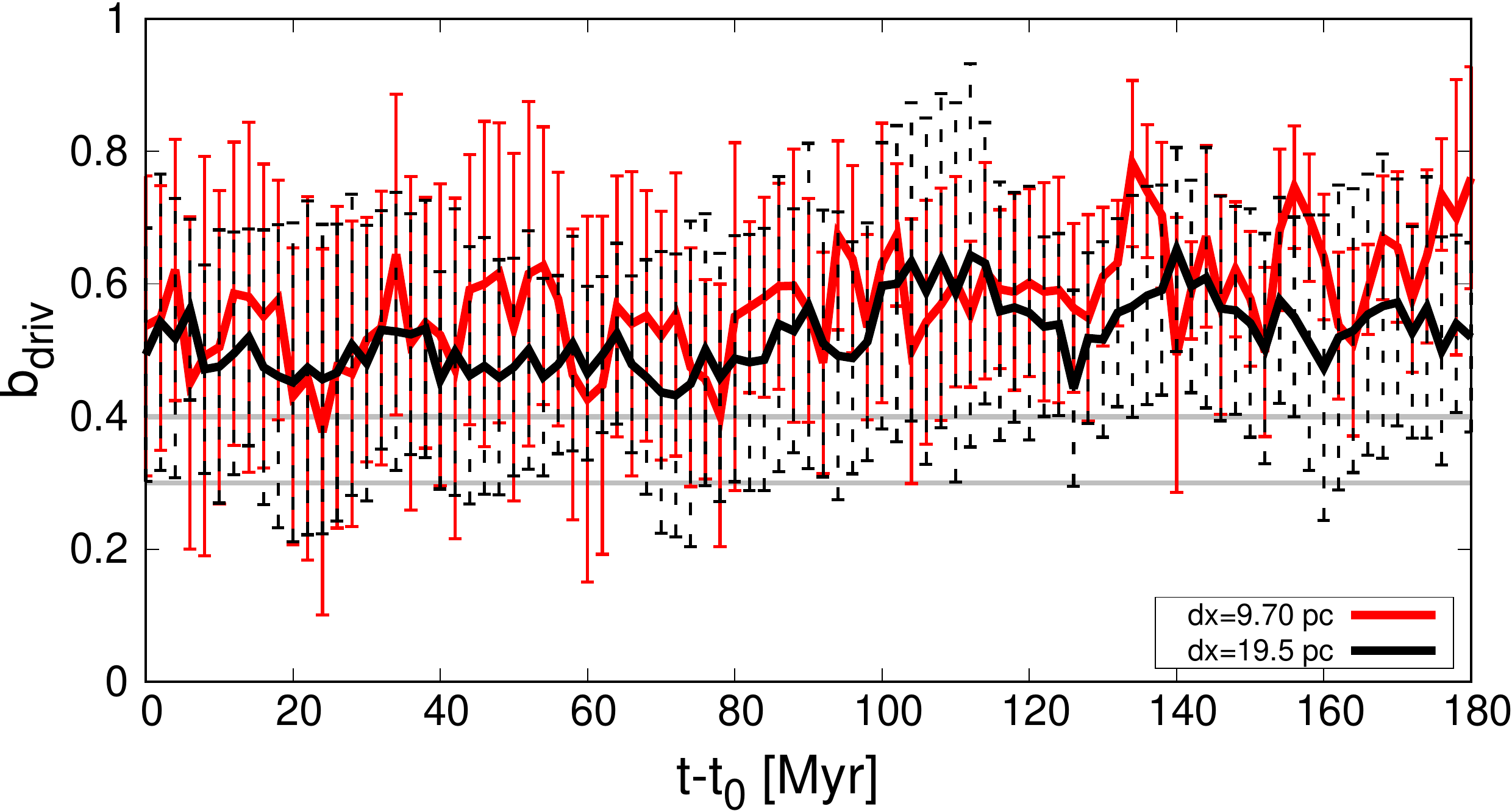}
    \caption{Resolution study of the evolution of the turbulence driving parameter for a period of 180\,Myr 
    starting at the identification of clouds ($t_0$). The error bars provide the standard deviation 
    $\Delta b_\mathrm{driv}$. 
    For the two resolutions, the driving parameter obtained from the Helmholtz decomposition appears to be largely converged. The grey horizontal lines denote purely solenoidal as well as natural driving.}
    \label{fig:app_resolstudy}
\end{figure*}

In Fig.~\ref{fig:app_resolstudy} we study the influence of the numerical resolution on the calculation of 
the average turbulence driving parameter, $b_\mathrm{driv}$. As investigated in 
\citet[][see also \cite{Koertgen17b}]{Jin17}, \ita{full} numerical convergence is achieved only with sub-parsec resolution. We show two resolutions, namely our fiducial 
resolution with $\Delta x=19.5\,\mathrm{pc}$ and a slightly higher one with $\Delta x=9.7\,\mathrm{pc}$. 
The time axis is given in time relative to the time when clouds are identified for the first time, $t_0$.\\
The driving parameter in the higher resolution run seems to fluctuate slightly more, but the very good 
agreement of the averages is convincing. Our explanation for this good correspondence is that we derive the 
driving parameter via the compressive ratio 
$\chi=\left<v_\mathrm{comp}^2\right>/\left<v_\mathrm{sol}^2\right>$, where resolution effects nearly cancel 
out.

\section{Notes on the Fourier decomposition}
\begin{figure}
    \centering
    \includegraphics[width=0.45\textwidth,angle=-90]{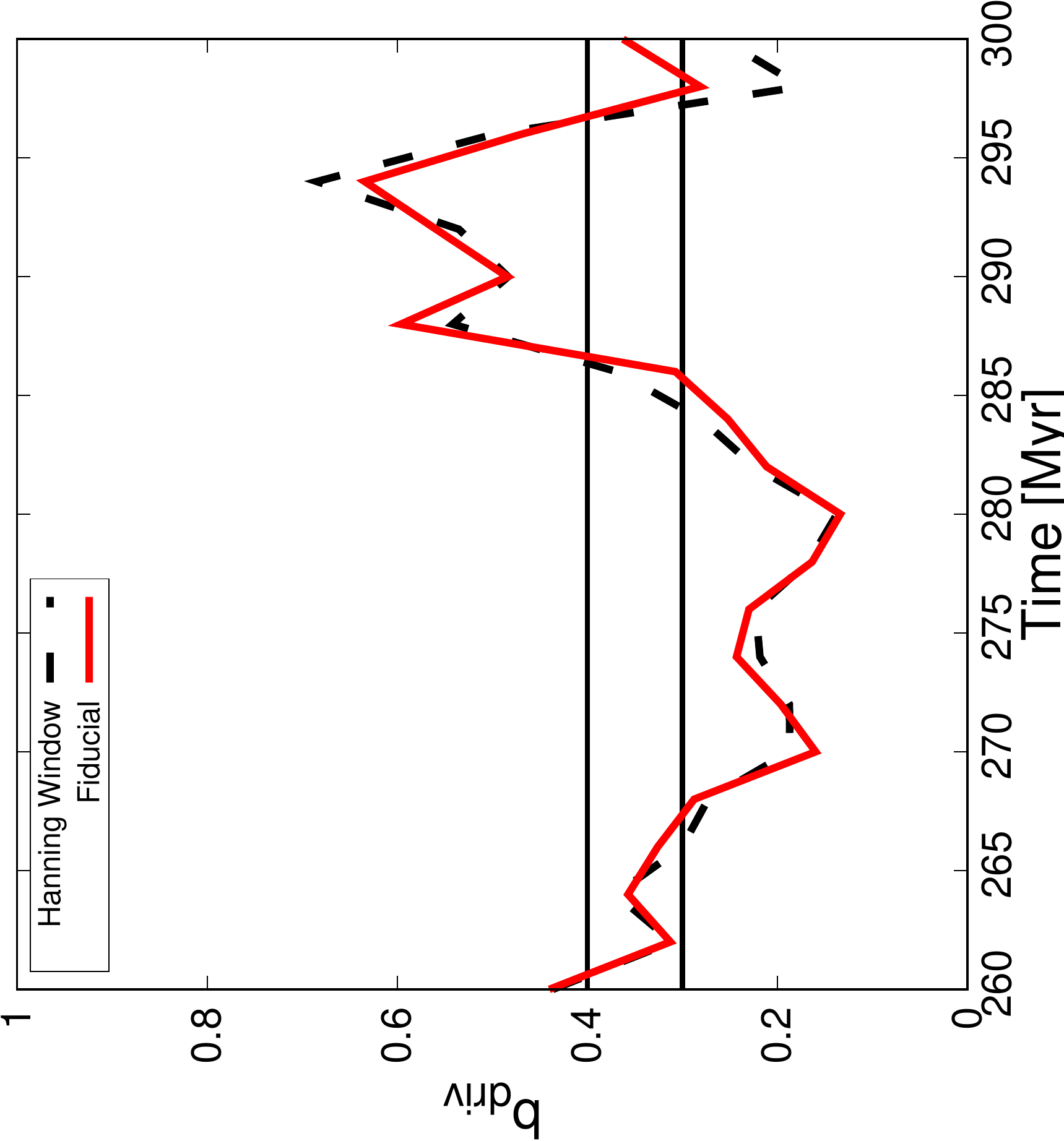}
    \caption{Comparison of the resulting driving parameter for the early phase of evolution of 
    the example cloud. The lines show the fiducial method (red) and the result after applying a 
    Hanning window to the turbulent velocity field. The agreement of the two methods ensures that 
    no numerical errors emerge from e.g. aliasing effects.}
    \label{fig:app_window}
\end{figure}

The extracted clouds in our simulations resemble regions with non-periodic boundary conditions. This 
can lead to significant errors in the resulting spectra due to aliasing effects. In such case, one 
commonly applies a window function, which enforces the field of interest to go smoothly to zero so 
that the box can be thought of as being periodic. The result of this \ita{windowing} is shown in 
Fig.~\ref{fig:app_window}. The modifications to the final driving parameter are only minor. This is 
due to the fact that it is obtained from the compressive ratio. Thus, possible errors will 
cancel out as they appear in both of the decomposed fields.

\end{appendix}

\section*{Data availability}
The data underlying this article will be shared on reasonable request to the corresponding author.
\bibliography{astro} 
\bibliographystyle{mn2e}
\end{document}